\documentclass[twocolumn,11pt]{article}
\pdfoutput=1
\usepackage[utf8]{inputenc}
\usepackage[english]{babel}
\usepackage[T1]{fontenc}
\usepackage{amsmath}
\usepackage{hyperref}
\usepackage{float}
\usepackage{tikz}
\usetikzlibrary{quantikz}
\usepackage{lipsum}

\begin{document}

\title{Group-invariant estimation of symmetric states generated by noisy quantum computers}

\author{Federico Holik$^{1}$, Marcelo Losada$^{2}$, Giannina Zerr$^{3}$, Lorena Rebón$^{1,\,4}$ and Diego Tielas$^{1,\,4}$}


\twocolumn[
  \begin{@twocolumnfalse}
    \maketitle
    \begin{center}

\begin{small}

\date{%
 1- Instituto de Física La Plata (CONICET-UNLP), Calle 113 entre 64 y 64 S/N, C.P. 1900, La Plata, Buenos Aires, Argentina \\
2- Facultad de Matemática, Astronomía, Física y Computación, Universidad Nacional de Córdoba, Av. Medina Allende S/N, Ciudad Universitaria, X5000, Córdoba, Argentina\\
3- Departamento de Física, Facultad de Ciencias Exactas y Naturales, Universidad de Buenos Aires, Pabellón 1, Ciudad Universitaria, C.P. 1428, Buenos Aires, Argentina\\
4- Departamento de Ciencias Básicas, Facultad de Ingeniería, Universidad Nacional de La Plata, Calle 48 y 116, C.P. 1900, La Plata, Buenos Aires, Argentina
}
\end{small}
\end{center}

    \begin{abstract}
The problem of quantum state estimation is crucial in the development of quantum technologies. In particular, the use of symmetric quantum states is useful in many relevant applications. In this work, we analyze the task of reconstructing the density matrices of symmetric quantum states generated by a quantum processor. For this purpose, we take advantage of an estimation technique that results to be equivalent to the quantum Maximum Entropy (MaxEnt) estimation, and which was recently adapted to quantum states with arbitrary symmetries.
The smart use of prior knowledge of the quantum state symmetries allows for a reduction in both, the number of measurements that need to be made on the system, and the size of the computational problem to store and process the data, resulting in a better overall performance of the estimator as well.
After performing numerical simulations, we implement some examples of symmetric states in IonQ quantum processors, and estimate them using the proposed technique. The results are in a good agreement with numerical simulations, showing that the proposed method is a good estimator that allows to save both, experimental and computational resources.
    \end{abstract}
\begin{small}
\centerline{\em Key words:Quantum State Estimation; Quantum Computers;
Group Invariant Quantum Tomography}
\end{small}

  \end{@twocolumnfalse}
]

\twocolumn

\section{Introduction}\label{s:Introduction}
In many situations,
it is necessary to determine which are the resources that a given device is able to produce. In fact, this is a relevant problem to solve in order to certify that such quantum device is reliable and secure.
Ideally, the most complete characterization of a quantum system that can be done is to perform a quantum state tomography,  given that, in principle, knowing accurately the quantum state actually produced would imply to know all possible resources at hand.
Notwithstanding, 
tomographic methods are 
hard to implement, since the size of the problem grows exponentially with the number of qubits. This complexity manifests itself both in terms of the number of measurements involved and in the computational complexity associated with the statistics processing as well.
In this context, it is essential to develop techniques that allow minimizing the number of measurements and the calculation time consumed by data post-processing. 

In previous works \cite{tielas2021performance,Corte2023}, we have developed a quantum tomography method, called  \textit{ group-invariant quantum tomography} (GIT). Through the use
of prior knowledge of the symmetry of an otherwise unknown state, we have found a parameterization of its density matrix which is specially convenient for quantum state estimation purposes.
This allows reducing the computational and experimental costs in many examples of interest, extending previous restricted versions of the problem \cite{PermutationallyInvariantQT} to arbitrary symmetries. As it is combined with MaxEnt and variational methods \cite{Goncalves-2013-MaxEntTomography}, in many cases, the number of measurements needed for a reasonable estimation can be reduced even further (see Ref.~\cite{tielas2021performance}, Section IV).

In this work, we apply the GIT technique to different symmetric states generated in both simulated and real quantum computers.
After reviewing the basic features of the GIT method in Section \ref{s:Overview}, we present, in Section \ref{s:Numerical_Simulations}, the results of several numerical simulations using the Amazon Braket SDK \cite{braket}.
Our results illustrate the differences between ideal behavior/performance and what should be expected in NISQ devices.
Moreover, we perform several experiments on different quantum computers in order to assess their capability of generating symmetric states. In Section \ref{s:Experimental_Results}, we present the results of these experiments, which are in good agreement with the numerical simulations.
Our findings suggest that GIT can be a relevant technique to certify quantum devices in the NISQ era. The main advantage is that it allows for a substantial reduction in the resources needed to estimate to what degree the devise was capable of generating a desired symmetric state. In Section \ref{s:Conclusions} we present our final conclusions and outline future work.

\section{Overview of GIT}\label{s:Overview}

In this section we give a brief overview of the GIT methodology. For details, see Ref.~\cite{tielas2021performance} and Ref.~\cite{Corte2023}.

The GIT methodology relies on a parametrization of a state which is known to have a certain symmetry (but unknown otherwise). The symmetry can be continuous or discrete. Given a symmetry $\mathcal{S}$, an orthogonal basis $\mathcal{B}_{\mathcal{S}}=\{S_{i}\}$ of Hermitian matrices associated to the symmetry is computed. This basis generates all possible symmetric states. Thus, an arbitrary density operator $\rho$ possessing the symmetry can be written as a linear combination

\begin{equation}\label{e:Parameterization}
    \rho = \sum^{r}_{i=1} \alpha_{i}S_{i}
\end{equation}

\noindent where the $\alpha_{i}$ are real parameters. The basis $\mathcal{B}_{\mathcal{S}}$ is computed numerically.
It is important to remark that the numerical computation can be performed even in cases where an exact formula for the parameterization of symmetric states is not known.
For many examples of symmetries, the number of elements of $\mathcal{B}_{\mathcal{S}}$ is much lower than that of the canonical basis of density operators. This implies two things:

\begin{itemize}\label{e:Parametrization}
    \item The number of experiments (measurement basis on which the statistics are performed) is substantially reduced.
    \item The size of the computational problem decreases significantly.
\end{itemize}

The above methodology allows to compute which observables give optimal information about the symmetric but otherwise unknown quantum state. Using the above parametrization, the following optimization problem is performed.

\begin{align}\label{e:VQT_GIT}
     \min_{\rho,\,\Delta} \left( \alpha\sum_{i\in\mathcal{I}} \Delta_{i} + \beta\sum_{i\notin \mathcal{I}}\mbox{tr}(E_{i}\rho)- \gamma \log\left(\det(\rho)\right)\right), \\ \nonumber
    \mbox{subject to}\,\,| \mbox{tr}(E_{i}\rho)-f_{i}|\leq \Delta_{i}f_{i} \,\,\, i \in\mathcal{I},\\\nonumber
    \Delta_{i}\geq 0,\\\nonumber
    \mbox{tr}(\rho)=1,\\\nonumber
    \rho \succeq 0,
\end{align}

\noindent with $\alpha,\beta,\gamma\geq 0$ to be adjusted in applications. The above optimization method is called Variational Quantum Tomography (VQT). In Ref.~\cite{Goncalves-2013-MaxEntTomography} it is proved that, up to four qubits, the above optimization (with $\alpha = \beta = 1$ and $\gamma = 0$) yields results that are in high agreement with those of the MaxEnt methodology. Some arguments are given to suggest that the VQT method agrees with MaxEnt for higher dimensions. The advantage of solving the above optimization instead of MaxEnt, is that the numerical problem becomes much softer. Thus, the combination of the VQT estimation with the use of the symmetric parametrization gives place to a powerful technique, which can be used to save a lot of computational an experimental resources in practice. In this work, we have empirically found that using different values for $\alpha$ and $\beta$, and using $\gamma \neq 0$, gives place to a better performance of the method. In particular, the term $\log(\det(\rho))$ improves the performance of the technique for the case of mixed states. This is suitable for dealing with noisy devices.

The observables used in practice need not be restricted to projection operators (neither POVMs). If the mean values of more general observables are used (as, for example, those associated to Pauli operators), the quantities $f_{i}$ might acquire negative values. In that case, one should do the replacement

$$|\mbox{tr}(E_{i}\rho)-f_{i}|\leq \Delta_{i}f_{i} \longrightarrow | \mbox{tr}(E_{i}\rho)-f_{i}|\leq \Delta_{i}|f_{i}|.$$

\noindent In this work, we restrict to projection operators for simplicity of implementation in quantum processors. But it is important to keep in mind that nowadays platforms (such as AWS Braket) allow to submit instructions for measuring general observables in an automated way. This is important for us, since the optimal observables for Werner states are not projections (see Ref.~\cite{Corte2023}), a subject that we will deal in future works.

Three very important remarks are in order:

\begin{itemize}
    \item There are techniques which are specifically adapted to permutationally invariant states (see Ref.~\cite{PermutationallyInvariantQT}). For that case, the GIT methodology yields equivalent results but, during the optimization process, the parametrization of the state is computed following a very different methodology. Thus, the analysis performed in this work both, in numerical simulations and actual quantum processors, is a test of the GIT method, and not of the so called-permutationally invariant quantum tomography.
    \item The GIT methodology can be applied numerically to families of states for which there is no known exact solution for the parametrization problem. For example, this is the case of Werner states. Notice that the parametrization used in previous methods, such as the Permutationally Invariant Quantum Tomography (PIT) \cite{PermutationallyInvariantQT}, are not well suited for Werner states. Parametrizations are known up to three qubits systems \cite{Eggeling2001}, and only abstract formulas are known for higher dimensions \cite{Eggeling_PhD}. The GIT methodology works for arbitrary symmetries, and does not depend on any exact parametrization.
    \item As is the case with MaxEnt, the method \ref{e:VQT_GIT} is expected to converge even when the number of measurements is below the quorum determined by the number of parameters occurring in Eq. \eqref{e:Parametrization}. As shown in Ref.~\cite{tielas2021performance} and Ref.~\cite{Goncalves-2013-MaxEntTomography}, for many quantum states, the methodology yields reasonable results even when the number of measurements is far from quorum. This feature allows to further reduce the resources needed to perform a reasonable estimation in many cases of interest.
\end{itemize}

\section{Numerical simulations}\label{s:Numerical_Simulations}

In this section we perform numerical simulations that illustrate the performance of the GIT methodology in different noisy quantum computing scenarios. We focus on amplitude damping, bit flip and depolarizing noise separately. We also vary the number of shots of the simulated statistics for different symmetric states. By contrasting the ideal case vs the noisy scenarios, we obtain a map which allows to establish in which regimes the GIT method is expected to yield relevant information about the system under study.

\subsection{Different shot numbers and noise levels}

Here we illustrate how the method behaves with regard to different number of shots and noise levels. With that aim, we devised the following numerical experiment.

\begin{itemize}
\item First, we choose some symmetric target states $\rho_{T}$, namely GHZ states for different qubit numbers and two-qubit Werner states\footnote{For a theoretical description of all the symmetric states used in this work, see Section \ref{s:Experimental_Results}.}.
\item Next, we generate synthetic statistical data in a quantum computer simulator with the Braket SDK \cite{braket}, using different levels of noise and shots numbers.
\item For a given number of shots, type and intensity of noise, we obtain $\rho_{R}$ (the actually generated state), $\rho_{GIT}$ (the state estimated using GIT), and $\rho_{cVQT}$ (the state estimated using the complete variational tomography).
\item Each simulation consists in preparing a noisy GHZ
or Werner state, simulating its statistics, and performing an estimation from the obtained data. This process is repeated $30$ times (thus, $30$ density operators are obtained). The repetition is done to gain knowledge about how robust the method is with regard to statistical fluctuations in data.
\item Finally, we compare $\rho_{R}$, $\rho_{GIT}$, and $\rho_{cVQT}$, to assess how close is the GIT estimation with regard to the real state and how much information is lost with regard to a complete tomographic estimation. We also compare the results with respect to $\rho_{T}$, the target state, which is the expected state that an ideal (noiseless) device would produce.
\end{itemize}

\begin{figure}[H]
\centering
\includegraphics[width=1\linewidth]{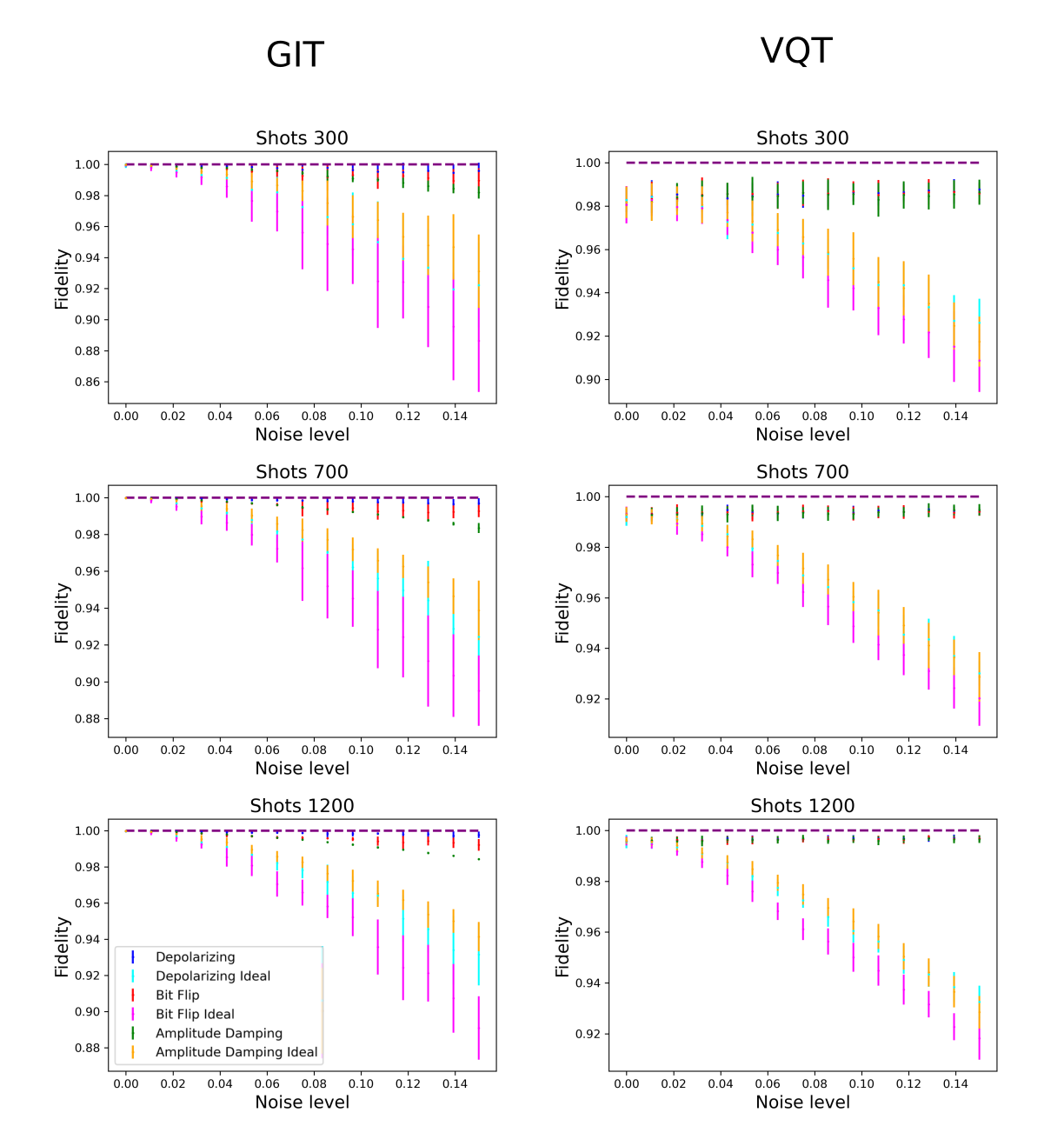}
\caption{Fidelities for GIT (left) and complete VQT (right) estimations obtained from the simulated statistics of a two-qubit Werner state, with $p=0.51$ (see Eqn. \eqref{e:Werner_2q_equation} below), with respect to the target and real state. Each tomography is repeated $30$ times (error bars). The fidelities are plotted as a function of the noise level for each type of noise and for different number of shots. Here and in the rest of the figures the fidelities with respect to the target state $\rho_{T}$ are referred as \textit{ideal}.}
\label{f:complete_vs_GIT_2q_werner}
\end{figure}

\begin{figure}[H]
\centering
\includegraphics[width=1\linewidth]{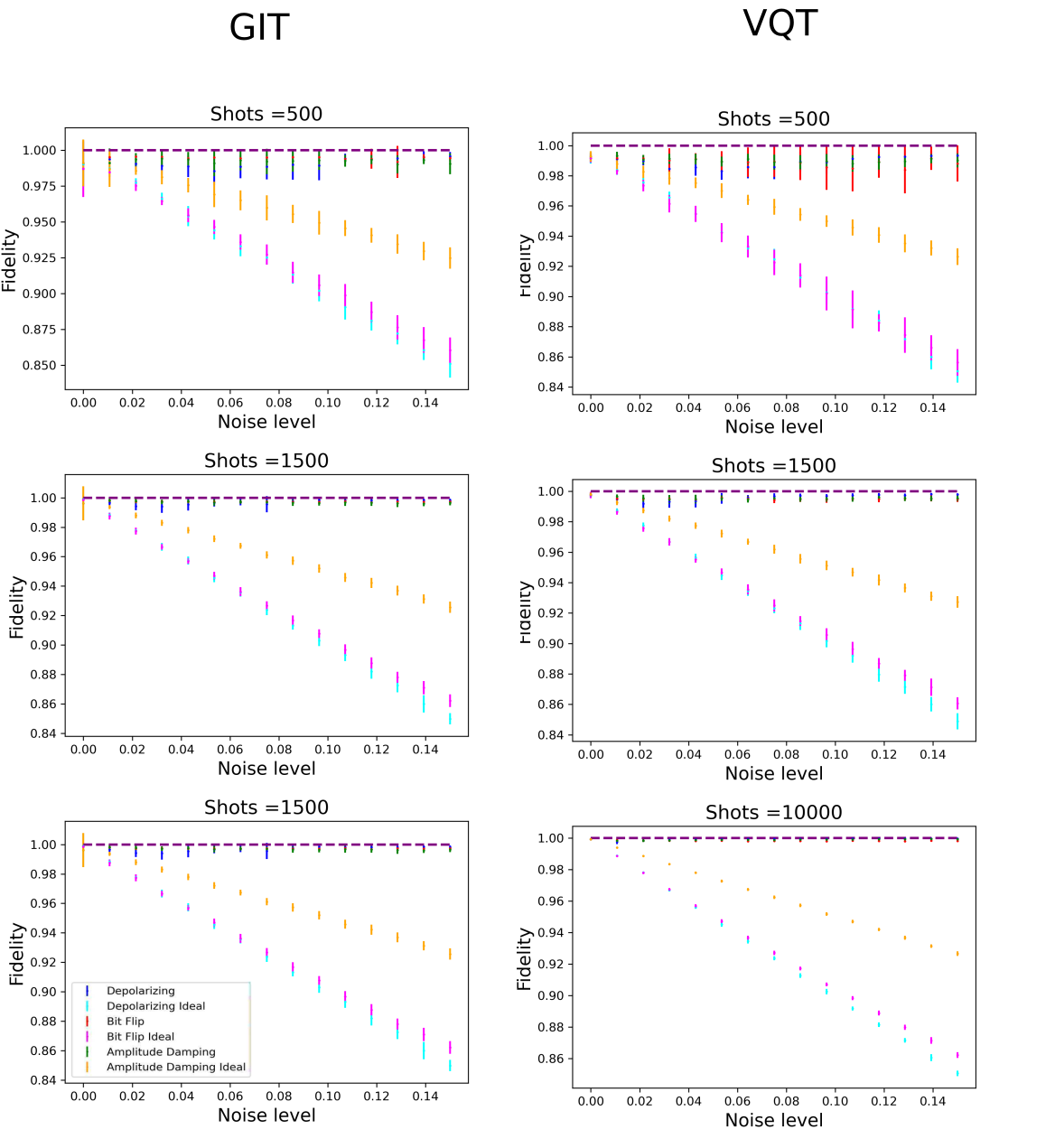}
\caption{Fidelities for GIT (left) and complete VQT (right) estimations obtained from the simulated statistics of a two-qubit GHZ state with respect to the target and real state. Each tomography is repeated $30$ times (error bar). The fidelities are plotted as a function of the noise level for each type of noise and for different number of shots.}
\label{f:complete_vs_GIT_2q_GHZ}
\end{figure}

\begin{figure}[H]
\centering
\includegraphics[width=1\linewidth]{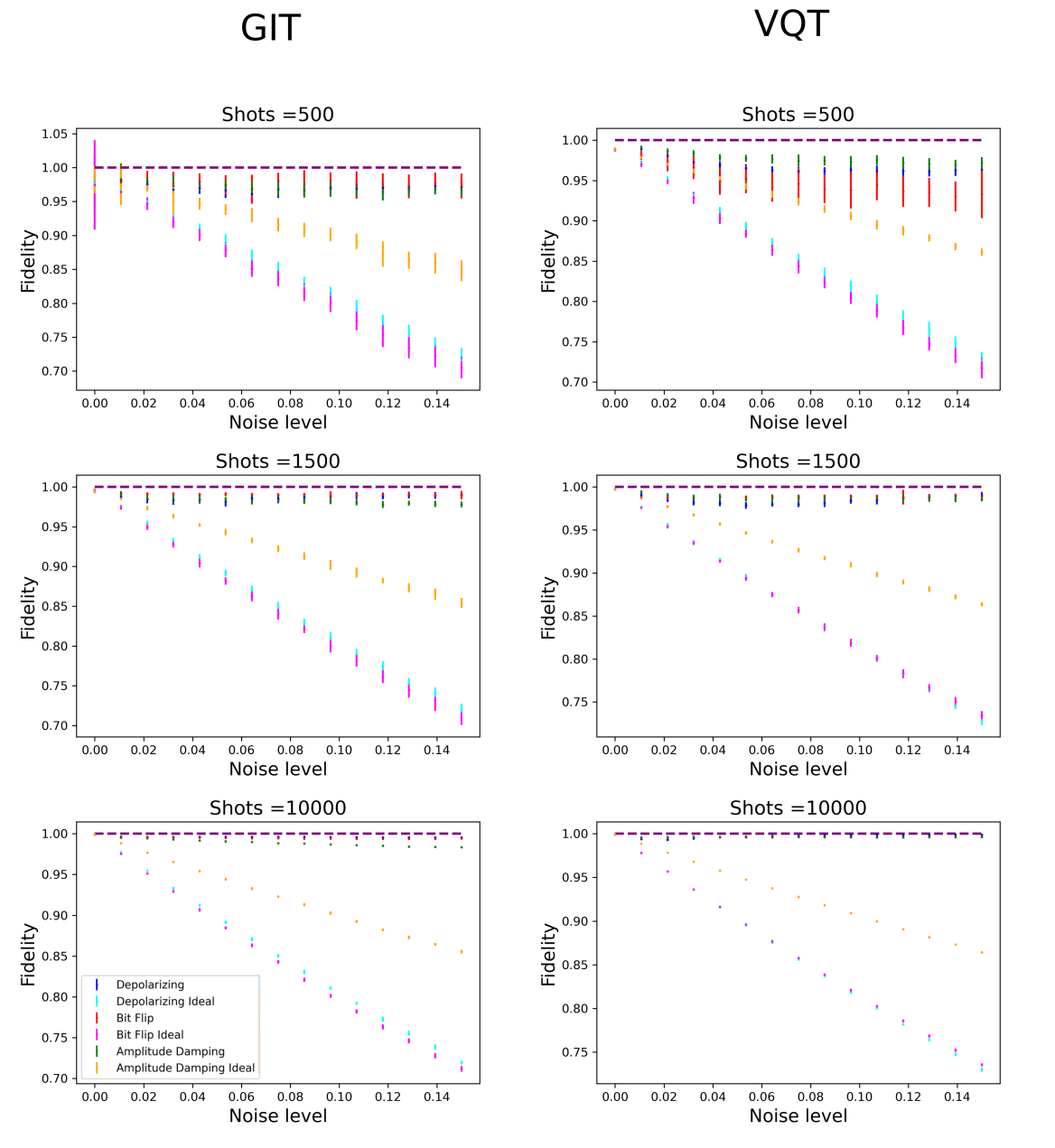}
\caption{Fidelities for GIT (left) and complete VQT (right) estimations obtained from the simulated statistics of a three-qubit GHZ state with respect to the target and real state. Each tomography is repeated $30$ times (error bar). The fidelities are plotted as a function of the noise level for each type of noise and for different number of shots.}
\label{f:complete_vs_GIT_3q_GHZ}
\end{figure}

\begin{figure*}
\centering
\includegraphics[width=1\linewidth]{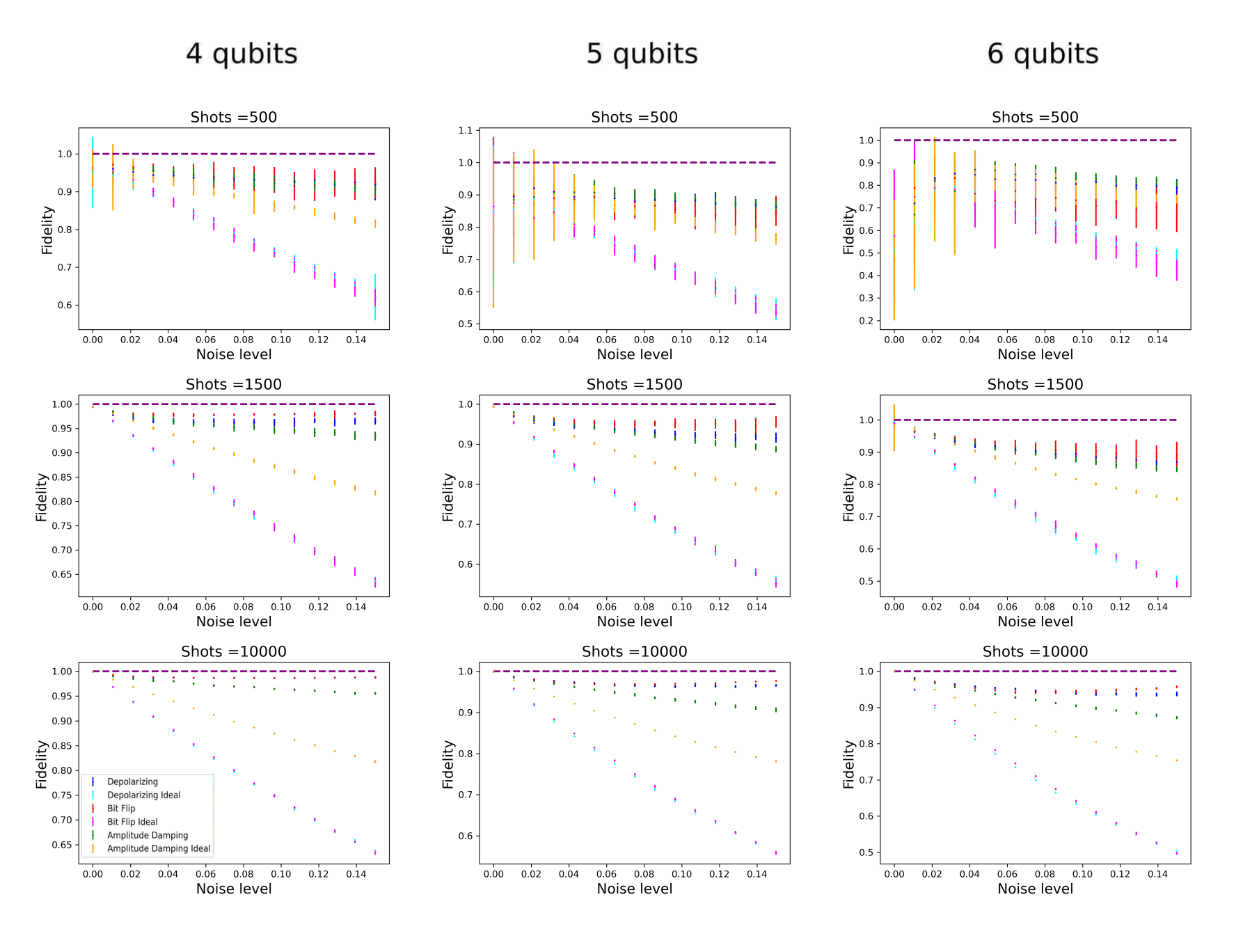}
\caption{Fidelities for GIT estimation obtained from the simulated statistics of four, five, and six-qubit GHZ states with respect to the target and real state. Each tomography is repeated $30$ times (error bar). The fidelities are plotted as a function of the noise level for each type of noise and for different number of shots.}
\label{f:GHZ_6q_permutational}
\end{figure*}

\begin{figure}[H]
\centering
\includegraphics[width=1\linewidth]{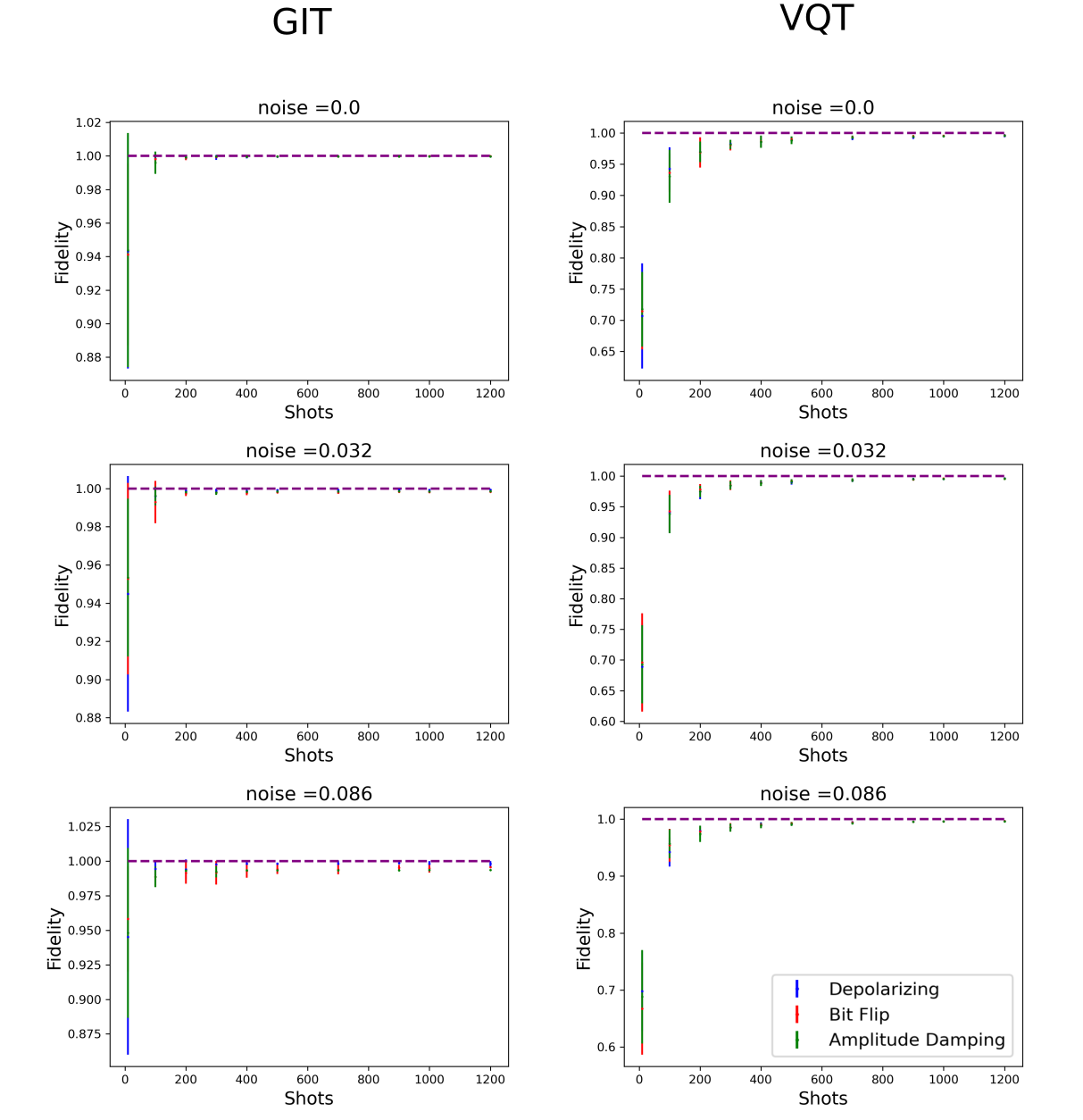}
\caption{Fidelities of GIT (left) and complete VQT (right) estimations obtained from the simulated statistics of a two-qubit Werner state, with $p=0.51$, with respect to the target and real state. Each tomography is repeated $30$ times (error bar). The fidelities are plotted as a function of the number of shots for different noise level for each type of noise.}
\label{f:shots_2q_Werner}
\end{figure}

\begin{figure}[H]
\centering
\includegraphics[width=1\linewidth]{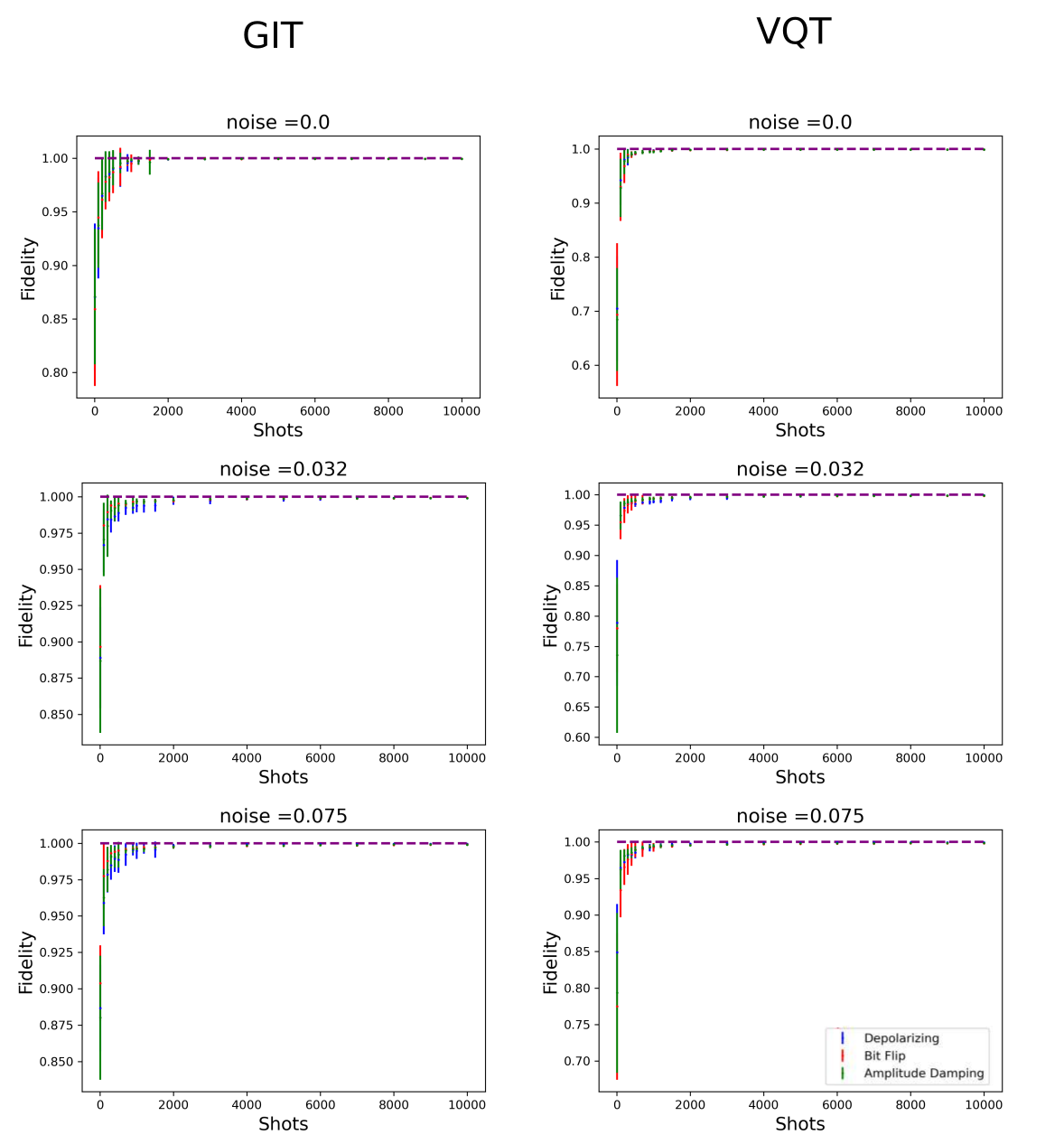}
\caption{Fidelities for GIT (left) and complete VQT (right) estimations obtained from the simulated statistics of a two-qubit GHZ state with respect to the target and real state. Each tomography is repeated $30$ times (error bar). The fidelities are plotted as a function of the number of shots for different noise level for each type of noise.}
\label{f:shots_2q_GHZ}
\end{figure}

\begin{figure}[H]
\centering
\includegraphics[width=1\linewidth]{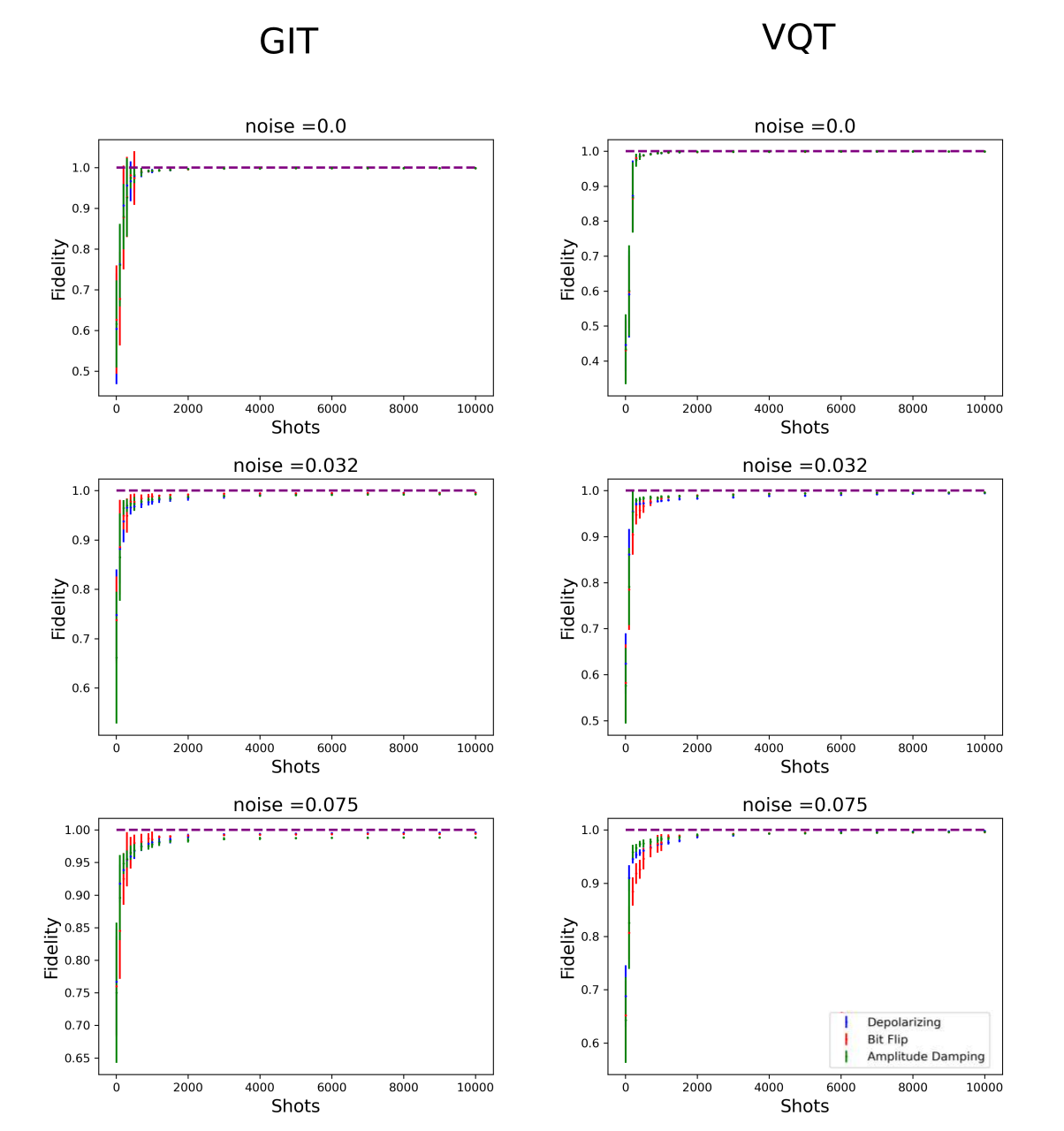}
\caption{
Fidelities for GIT (left) and complete VQT (right) estimations obtained from the simulated statistics of a three-qubit GHZ state with respect to the real state. Each tomography is repeated $30$ times (error bar).
The fidelities are plotted as a function of the number of shots for different noise level for each type of noise.
We observe that, in the GIT method, the fluctuation in the results is high when the number of shots falls short.}
\label{f:shots_3q_GHZ}
\end{figure}

\begin{figure*}
\centering
\includegraphics[width=0.9\linewidth]{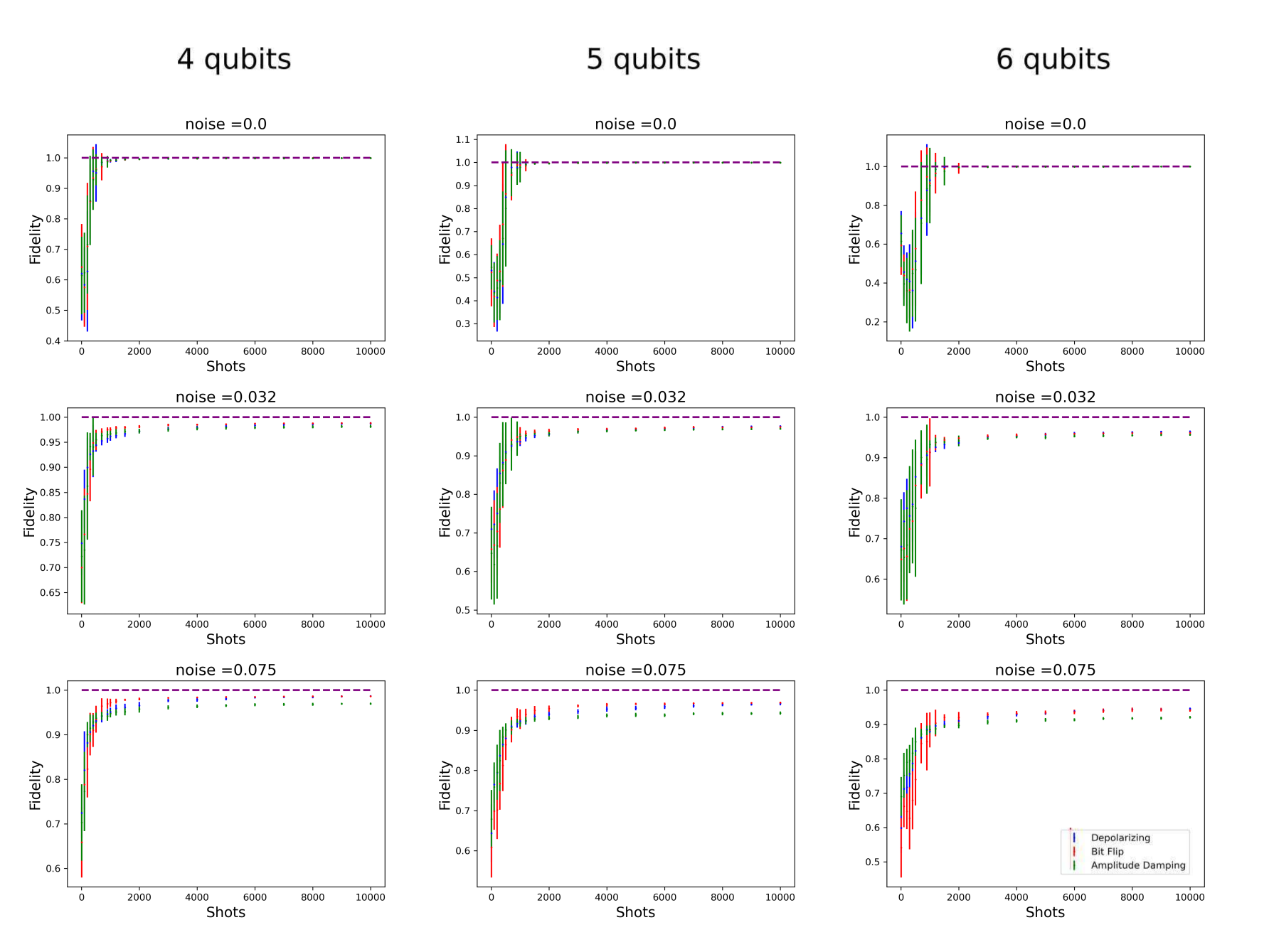}
\caption{Fidelities for GIT estimation obtained from the simulated statistics of four, five, and six-qubit GHZ state with respect to the target and real state. Each tomography is repeated $30$ times (error bar). The fidelities are plotted as a function of the number of shots for different noise levels for each type of noise.}
\label{f:shots_6q_GHZ}
\end{figure*}

The results obtained are presented in Figures \ref{f:complete_vs_GIT_2q_werner} to \ref{f:shots_6q_GHZ}. In Figures \ref{f:complete_vs_GIT_2q_werner}, \ref{f:complete_vs_GIT_2q_GHZ}, and \ref{f:complete_vs_GIT_3q_GHZ}, we show the results obtained for the simulated statistics of a two-qubit Werner state (with $p=0.51$) and $GHZ$ states of two and three qubits, respectively. As expected, when the number of shots increases, the error bars tend to decrease substantially.
But, for low number of shots, the dispersion is not negligible. Moreover, as noise increases, the fidelities of the estimated states $\rho_{GIT}$ and $\rho_{cVQT}$ with regard to the target state $\rho_{T}$ tend to decrease significantly\footnote{We refer to these fidelities as \textit{ideal}. For example, ``depolarizing ideal'' refers to the fidelity between the estimated state using data simulated with a depolarizing channel and the target state.}.
However, the fidelities obtained between
$\rho_{GIT}$ and $\rho_{R}$,
and
$\rho_{cVQT}$ and $\rho_{R}$
tend to stabilize in values above $98\%$.
The advantage of using GIT method is that it consumes much less resources than the complete VQT, giving an equivalent performance. Thus, under the hypothesis that the unknown state is symmetric (or approximately symmetric), the GIT method shows itself as very promising and reliable. Similar results are obtained for a higher number of qubits. In Figure \ref{f:GHZ_6q_permutational} we show the results of  GIT estimation for
four, five, and six-qubit GHZ states.
In these examples the error bars reach a stable behavior with a greater number of shots. Furthermore, the fidelities between
$\rho_{GIT}$ and $\rho_{R}$ tend to stabilize in different values for the different types of noise, being amplitude damping the more aggressive one. The relatively slight decrease in the performance is both, due to numerical errors during the optimization process, and the fact that noise alters the symmetry of the state. Both effects begin to play an important role when the number of qubits increases. Overall, the technique is very effective, since the balance between savings in computational resources and relatively small error rates makes the GIT method a good candidate for assessing quantum devices.


In Figures \ref{f:shots_2q_Werner}, \ref{f:shots_2q_GHZ}, \ref{f:shots_3q_GHZ} and \ref{f:shots_6q_GHZ}, we show different plots where, for a fixed noise level, the number of shots is increased. These plots illustrate how the performance of the methods behaves as a function of the number of shots. The results indicate that GIT and complete VQT methodologies are essentially equivalent, the main difference being that the GIT method yields highly fluctuating results when the number of shots stand below a certain threshold. This threshold depends on the number of qubits and must be taken into account if the GIT methodology is to be used.

\subsection{Summary}

The numerical simulations indicate that:

\begin{itemize}
\item In the ideal case, the quantum device should produce the target state. As expected, in that case, the GIT method yields the correct state estimation.
\item When noise is present, the fidelity between $\rho_{GIT}$ and $\rho_{T}$ tends to decrease. But still, the GIT method serves as a good estimator of the real state $\rho_{R}$ generated by the quantum device.
\item For reasonably low noise levels, the GIT method and the complete VQT yield equivalent results.
\item If the noise is high or the control fails significantly,
the GIT method will reveal a high divergence between the target and the estimated states. This can be used as a quick and inexpensive method to assess to which degree a given quantum device is able to produce the expected symmetric state. In this way, the performance of the GIT method is a reliable indicator of how good a device is from the point of view of the end user (preferable to full tomography in terms of resource savings).
\end{itemize}

\section{Implementation on quantum computers}\label{s:Experimental_Results}

In this section we show some implementations of the GIT and complete VQT techniques in Aria-1 and Harmony IonQ quantum processors \cite{ionq}. The results are in agreement with what was expected from the numerical simulations. In general, the fidelity between the estimated states obtained with GIT and complete VQT are high, giving an average of $0.96$ among all the performed experiments.
This indicates that, for the case of approximately symmetric states, the GIT methodology yields results which are essentially equivalent to that of the complete VQT (with the advantage of saving a lot of resources). For some cases, we have also performed a standard MaxLik estimation, and we have found that there is a good agreement between that technique and GIT.

\begin{figure*}
\centering
\includegraphics[width=1\linewidth]{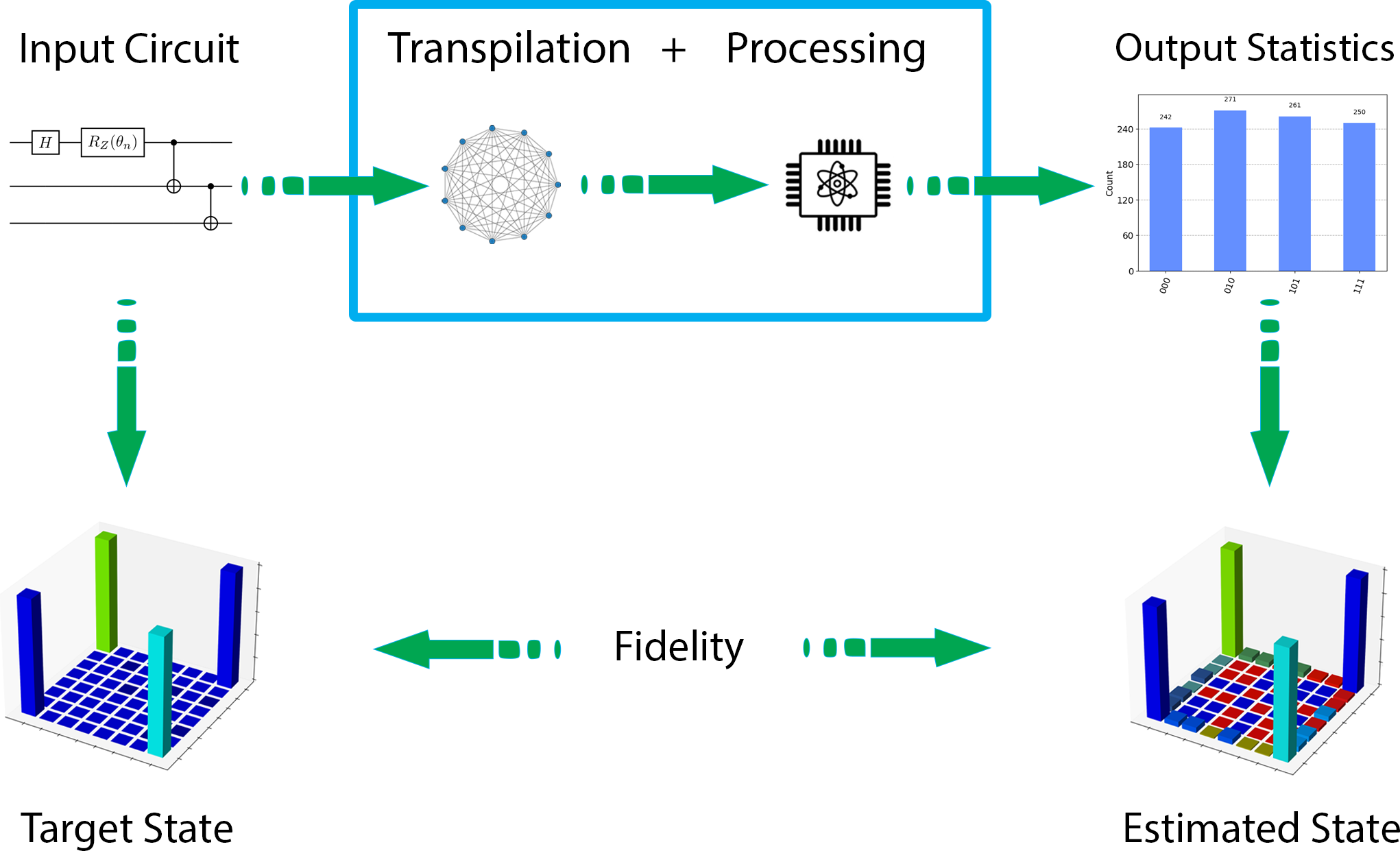}
\caption{
An input circuit is designed in some universal programming language. Each circuit contains a list of orders of a target state $\rho_{T}$ to be created, together with the instructions for a measurement in a certain basis. There is one circuit for each measurement basis. Next, each circuit associated to a target state and measurement basis is transpiled into native gates of the quantum hardware to be used, before it is sent to run to the actual processor. The output statistics of the different measurement basis is collected at the end of the process. Using the output statistics of all desired measurement basis, the estimation is performed on a classical hardware, giving place to $\rho_{GIT}$. At the end of the process we compare $\rho_{T}$ with $\rho_{GIT}$. A similar methodology is used for the rest of the estimation techniques, yielding $\rho_{cVQT}$ and $\rho_{MaxLik}$. It is crucial to have in mind that any possible deviations are due not only to the noisy processor, but also to any possible errors occurring during the transpilation process and numerical errors occurring when processing the output data. Thus, one can imagine anything that happens inside the blue box as a black box, whose behavior lies outside of the control of the end user.}
\label{f:experiment}
\end{figure*}

Each experiment is designed as follows. A target state $\rho_{T}$ is described as a set of instructions $I_{\rho_{T}}$ in some quantum computing programming language. The instructions $I_{\mathcal{B}_{i}}$ for measuring in a particular basis $\mathcal{B}_{i}$ are also added to the set of instructions. Thus, on each run of the experiment, we send to a quantum processor a set of instructions $I_{\rho_{T}}+I_{\mathcal{B}_{i}}$, and repeat them a certain number of times in order to obtain the ouput statistics for that basis. In the quantum computing jargon,
the set of instructions is called a \textit{task}
and the number of repetitions is called  number of \textit{shots} associated to that task. The AWS Braket API \cite{braket} is set in such a way that the set of instructions for a given task is first transpiled into native gates. Next, each circuit associated to a target state and measurement basis is transpiled into native gates of the quantum hardware to be used. The output statistics for different basis is collected and, out of them, a estimated state $\rho_{E}$ is obtained, where the subindex $E$ refers to the estimation method. Depending on the state, we use different types of estimation techniques, such as GIT, VQT, and MaxLik. The complete scheme is depicted in Figure \ref{f:experiment}. Notice that, by comparing the target state $\rho_{T}$ defined by the input set of instructions vs the estimated $\rho_{E}$ based on the output statistics, we are possibly including inefficiencies during the transpilation process. But this is relevant for the end user of the quantum device as a whole, since in everyday use a reliable automatic transpilation mechanism becomes a critical feature to consider. According to the simulations of the previous section, if the inaccuracies in the state preparation are not so rough, there should be a high degree of agreement between the target and estimated states. And, in many situations, even if the estimated state is not reached, one can trust that $\rho_{GIT}$ is a reasonable estimator of the actual state produced by the processor. Finally, we remark that measuring in a given basis implies the use of an extra set of instructions besides those specific of the state preparation $I_{\rho_{T}}$, which might become new sources of inaccuracies. This is a crucial point since, intuitively, acquiring more information should result in a more precise estimation of the unknown state. But, due to imperfections in the processor and transpilation process, acquiring more information might imply the counterintuitive effect of yielding more inaccurate results. The moral is that, sometimes, handling less information results in fewer errors. This is a point in which the GIT method represents an advantage with regard to other methods.

Another important aspect of our experiments is that, in most cases, we sent several states in parallel. Thus, for example, for a parametrized family of three-qubit states depending on a phase $\theta$, we can accommodate in Harmony three states with different values of $\theta$, since there are eleven qubits available in the processor. Since, in principle, the qubits associated to the different states do not interact, one can measure each state independently and, ideally, one should obtain the same results as by performing three different tasks, each one for each state. But the situation is different from the expected ideal case.

First of all, it must be taken into account that, if there are several qubits available, the transpiler will choose a sequence of native gates to obtain a global unitary to obtain the target state, and will subsequently implement the qubit reading procedure. By using as many qubits as possible to accommodate multiple states in a task, we are testing the processor in a more challenging context. By sending three three-qubit states to an eleven-qubit processor, we send a single global task of a nine-qubit state. Of course, this will stress the transfer and reading process.
%
%

\subsection{Different types of permutationally invariant states}

In this section we consider different types of permutationally invariant states. We start by considering states of the form
\begin{equation}\label{e:Parallel_phases}
|\psi_{\theta}\rangle = \frac{1}{\sqrt{2}}(|0...0\rangle + e^{i\theta} |1...1\rangle)
\end{equation}

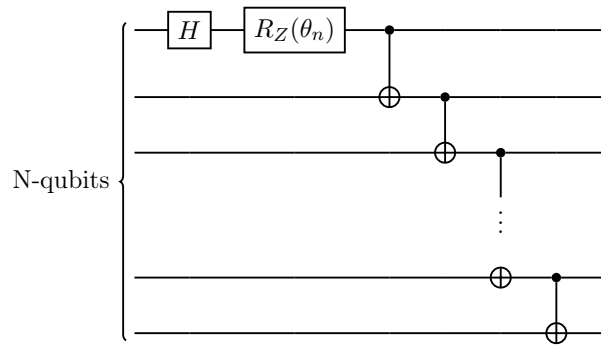
\begin{figure}[H]
\centering
 \resizebox{\columnwidth}{!}{
\begin{quantikz}
\lstick[6]{N-qubits}  & \gate{H} &\gate{R_Z(\theta_n)}& \ctrl{1} &\qw&\qw &\qw  &\qw\\
&\qw&\qw&  \targ{}  &\ctrl{1} & \qw & \qw   &\qw      \\
&\qw&\qw&  \qw & \targ{} & \ctrl{1} & \qw& \qw \\
&&&&& \vdots  && \\
&\qw&\qw&  \qw & \qw&  \targ{}  & \ctrl{1}&\qw \\
&\qw&\qw&  \qw & \qw & \qw & \targ{}  & \qw\\
\end{quantikz}}
\caption{
Scheme of the circuits used to generated
instances of the permutationally invariant states given in Eq. \eqref{e:Parallel_phases}, for different values of $\theta_{n}$.}
\label{f:Circuit_7q_GHZ_parallel}
\end{figure}

\noindent
By varying the phase of the circuit, one can obtains different permutationally invariant states.
We implement the above states for systems of two, three and seven qubits for different values of the phase parameter $\theta$. The circuit is displayed in Figure \ref{f:Circuit_7q_GHZ_parallel}.
%

\begin{figure}[H]
\centering
\includegraphics[width=1\linewidth]{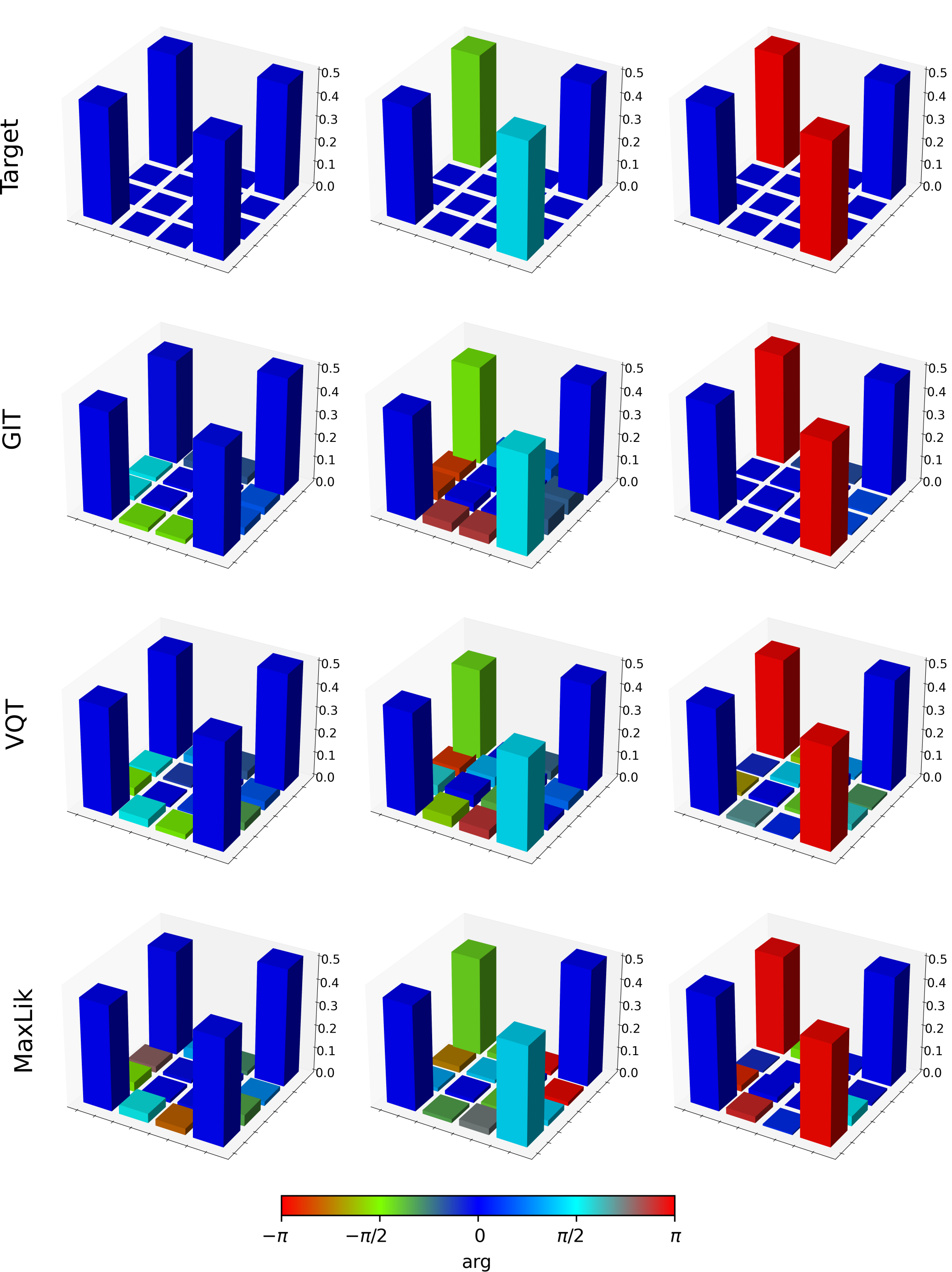}
\caption{
Three instances of two-qubit states of the form of Eq. \eqref{e:Parallel_phases} generated in Aria-1 quantum computer,
using $500$ shots for each measurement basis. From left to right, the angles are given by $\theta=0$, $\theta=5\frac{\pi}{22}$, and $\theta=\frac{\pi}{2}$.
From top to bottom, we show the target state, and the estimated states using GIT, complete VQT, and MaxLik methods.
For twelve expriments, the mean fidelities obtained with the different estimation methods are:
GIT vs Target : $0.94$,
cVQT vs Target: $0.93$,
MaxLik vs Target: $0.94$,
GIT vs cVQT : $0.96$,
GIT vs MaxLik: $0.97$,
cVQT vs MaxLik: $0.98$.
The fidelities indicate a good level of agreement between the different techniques.}
\label{f:Aria-1-2q-Parallel-Phases-07-06-2023}
\end{figure}

\begin{figure}[H]
\centering
\includegraphics[width=1\linewidth]{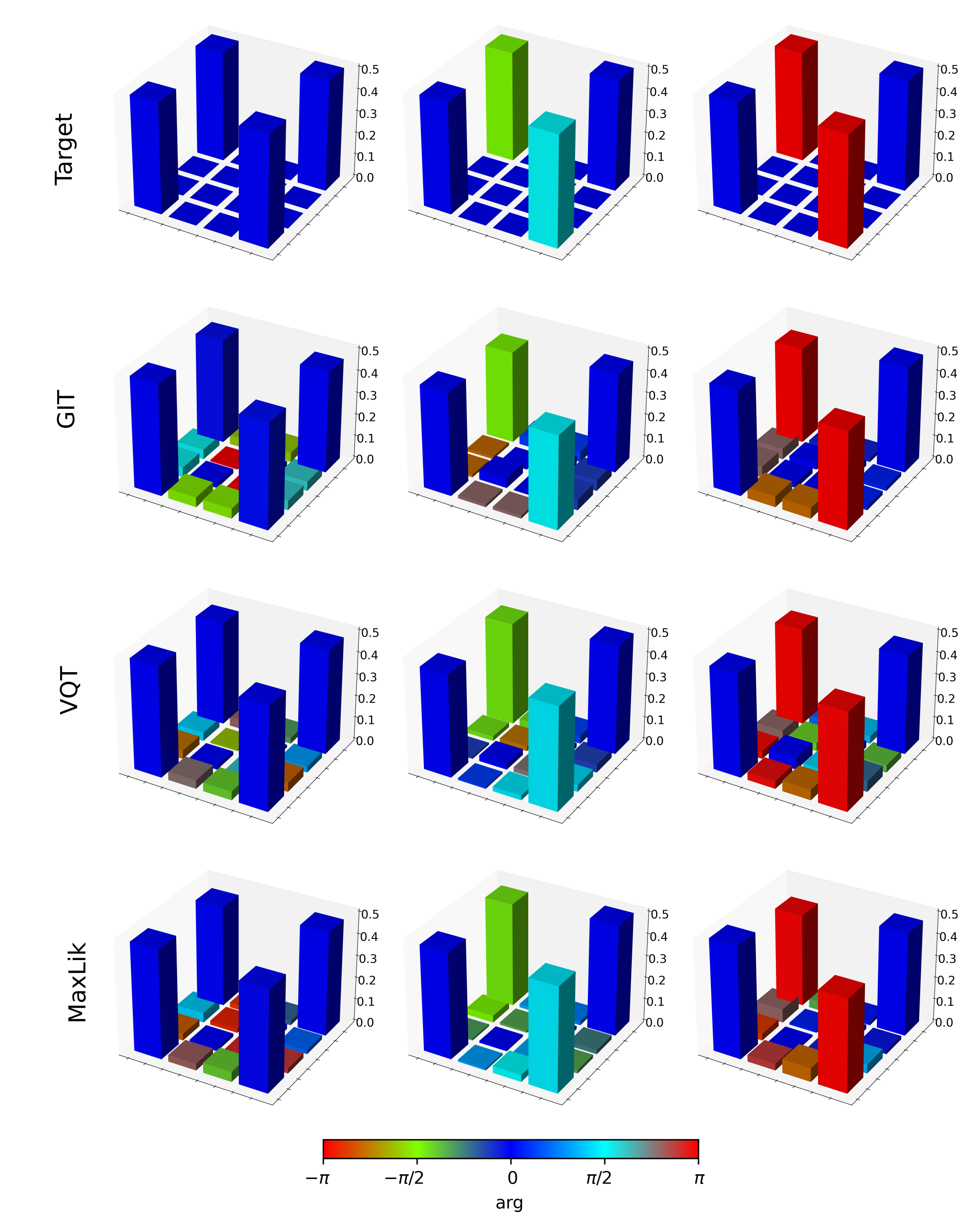}
\caption{Three instances of two-qubit states of the form of Eq. \eqref{e:Parallel_phases} generated in
Harmony quantum computer, using $500$ shots for each measurement basis. From left to right, the angles are given by $\theta=0$, $\theta=2\frac{\pi}{8}$, and $\theta=\frac{\pi}{2}$.
From top to bottom, we show the target state, and the estimated states using GIT, complete VQT, and MaxLik methods.
The mean fidelities obtained with the different estimation methods are:
GIT vs Target: $0.95$,
cVQT vs Target: $0.96$,
MaxLik vs Target: $0.96$,
GIT vs cVQT: $0,96$,
GIT vs MaxLik: $0.97$,
cVQT vs MaxLik: $0.96$.
The fidelities indicate a good level of agreement between both techniques.}
\label{f:Harmony-1-2q-Parallel-Phases-07-06-2023}
\end{figure}

\begin{figure}[H]
\centering
\includegraphics[scale=0.3]{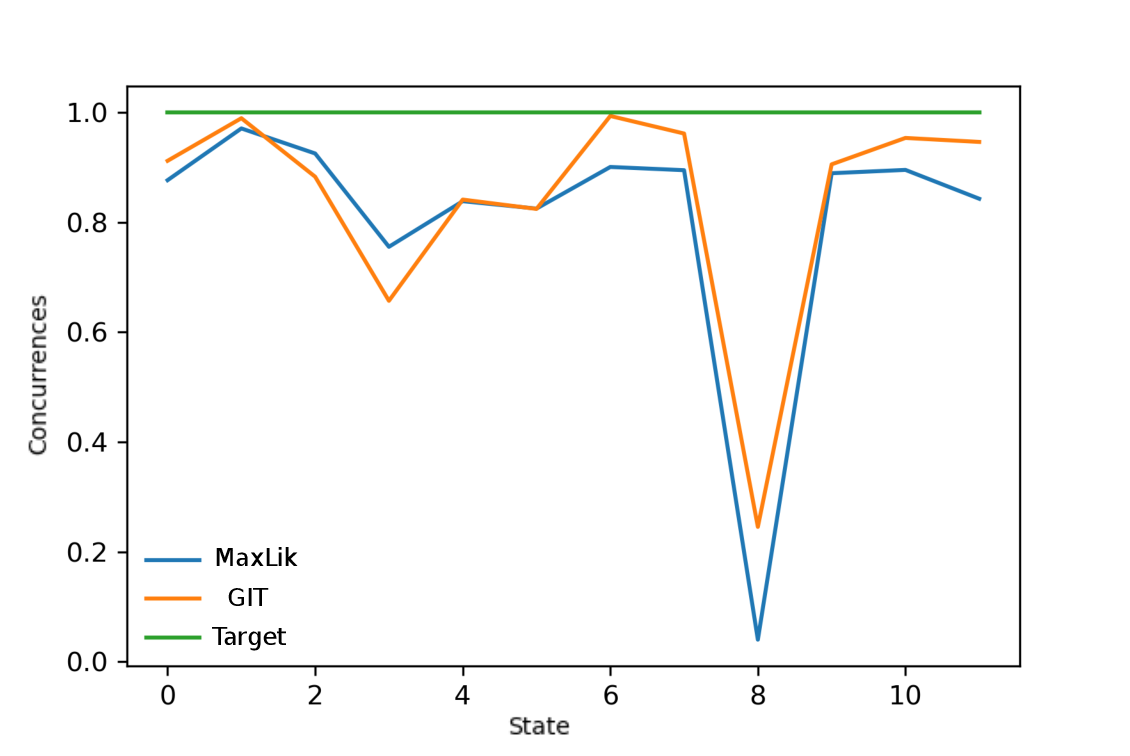}
\caption{Concurrences of all the two-qubit states of the form of Eq. \eqref{e:Parallel_phases} which were generated on the Harmomy processor (twelve in total; some of them are displayed in Figure \ref{f:Aria-1-2q-Parallel-Phases-07-06-2023}). The angles are given by $\theta_{j}=j\frac{\pi}{22}$, with $j=1,...,11$. The order in which the states are displayed is not relevant. The mean values of the absolute value of the differences between GIT and Target, GIT and MaxLik, and MaxLik and Target are $0.16$, $0.06$ and $0.20$
, respectively. Notice that, even for the ninth state, where the processor failed to produce the target state, there is still reasonable agreement between the results obtained using GIT and MaxLik.}
\label{f:2q_GHZ_concurrences_07_06}
\end{figure}

\begin{figure}[H]
\centering
\includegraphics[width=1\linewidth]{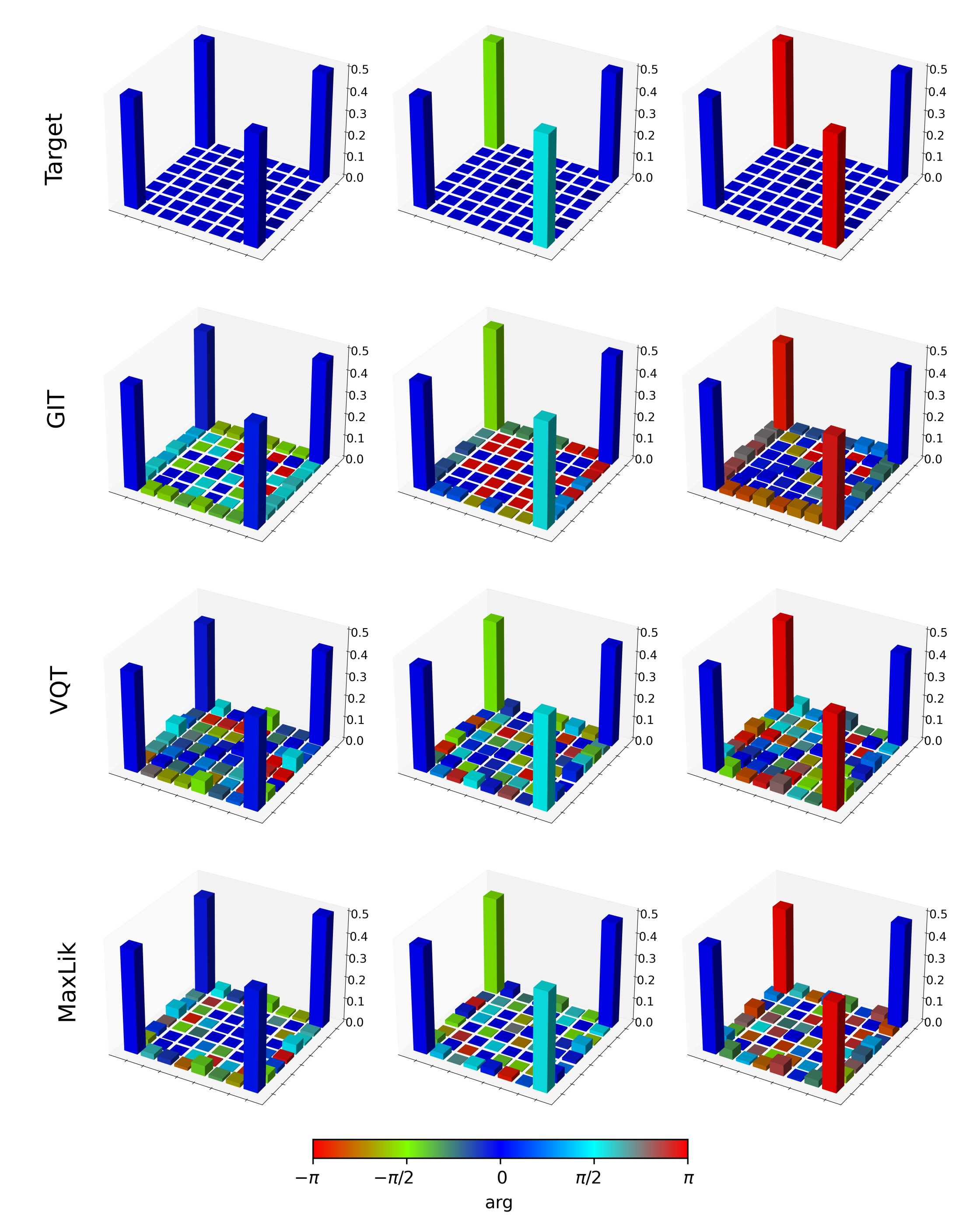}
\caption{Three instances of three-qubit states of the form of Eq. \eqref{e:Parallel_phases} generated in  Harmony quantum computer, using $1000$ shots for each measurement basis.
From left to right, the angles are given by $\theta=0$, $\theta=\frac{\pi}{4}$, and $\theta=\frac{\pi}{2}$.
From top to bottom, we show the target state, and the estimated states using GIT, complete VQT, and MaxLik methods.
The mean fidelities obtained with the different estimation methods are:
GIT vs Target: $0.96$,
cVQT vs Target:  $0.90$,
MaxLik vs Target: $0.96$,
GIT vs cVQT: $0.90$,
GIT vs MaxLik:  $0.97$,
cVQT vs MaxLik: $0.96$.
The fidelities indicate a good level of agreement between the different techniques.}
\label{f:3q_GHZ_parallel}
\end{figure}

\begin{figure}[H]
\centering
\includegraphics[width=1\linewidth]{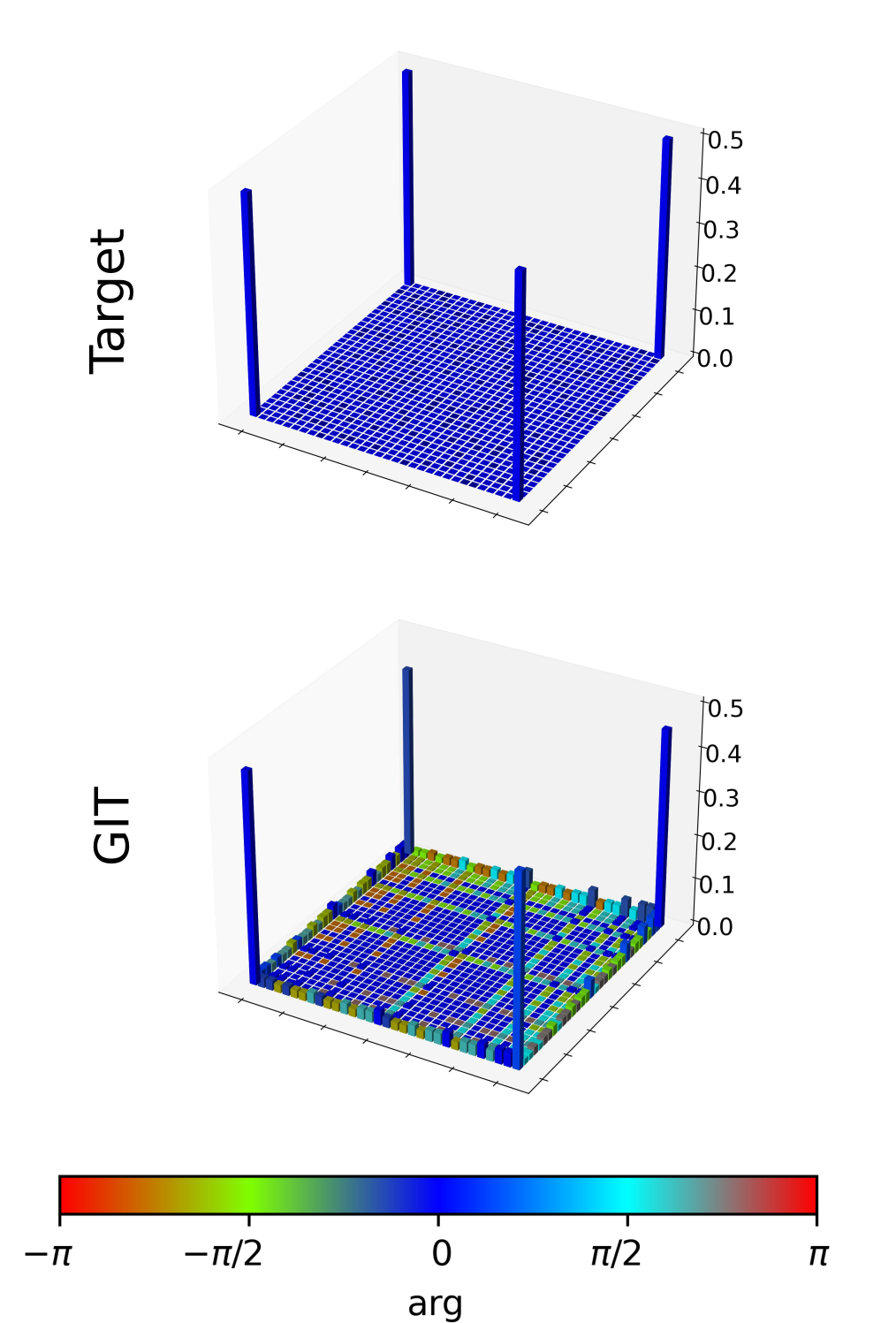}
\caption{Target state and GIT estimation of a five-qubit GHZ state generated in Harmony quantum computer, using $1500$ shots for each measurement basis. The obtained fidelity between both states is $0.92$.}
\label{f:5q_GHZ_Harmony}
\end{figure}

\begin{figure}[H]
\centering
\includegraphics[width=1\linewidth]{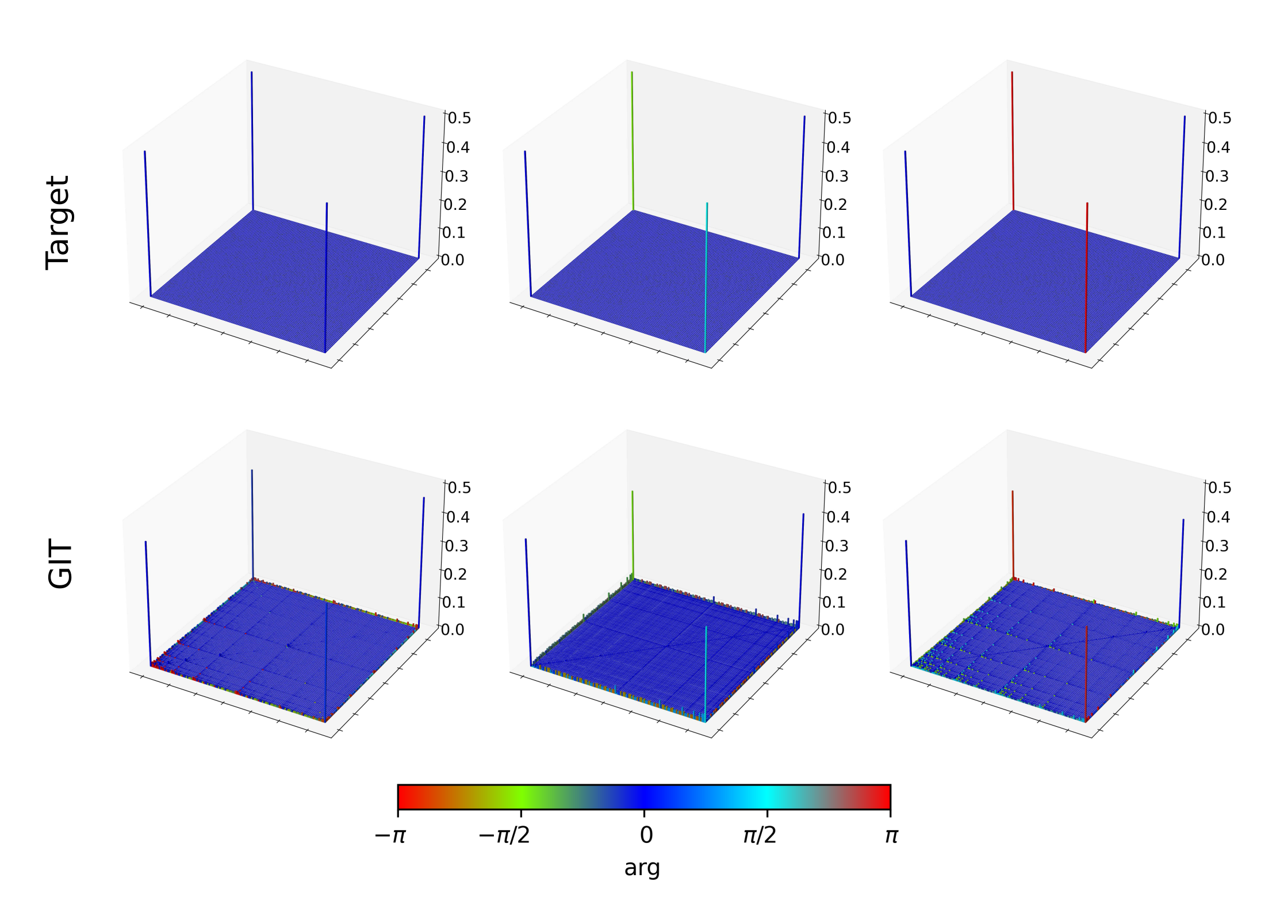}
\caption{Three instances of seven-qubit states of the form of Eq. \eqref{e:Parallel_phases} generated in Aria-1 quantum computer,
using $4000$ shots for each measurement basis.
From left to right, the angles are given by $\theta = 0$, $\theta = \frac{\pi}{2}$ and $\theta = \pi$.
From top to bottom, we show the target state, and the estimated state using GIT method.
The fidelities between the GIT estimation and the target state are given by $0.91$, $0.87$, and $0.86$, 
respectively.}
\label{f:7q_GHZ_Aria-1}
\end{figure}

In Figures \ref{f:Aria-1-2q-Parallel-Phases-07-06-2023} and \ref{f:Harmony-1-2q-Parallel-Phases-07-06-2023} we show the results obtained for some instances of two-qubit states of the form given in Eq. \eqref{e:Parallel_phases}, performed on Aria-1 and Harmony processors, respectively. All the plots of density matrices are performed using QuTiP \cite{qutip1, qutip2}. The fidelities between $\rho_{T}$ and $\rho_{GIT}$ are quite high. For the case of two and three qubits, we have also implemented the measurements needed for a complete estimation of the state, in order to perform a comparison with $\rho_{cVQT}$ and $\rho_{MaxLik}$. In Figure \ref{f:2q_GHZ_concurrences_07_06} we show the concurrences
of $\rho_{T}$, $\rho_{cVQT}$ and $\rho_{MaxLik}$ for all experiments that we performed on the Harmony processor (twelve in total). In Figures \ref{f:3q_GHZ_parallel},
\ref{f:5q_GHZ_Harmony} and \ref{f:7q_GHZ_Aria-1}, we show the results obtained for three, five and seven qubits, respectively.

For the two-qubit case, the basis for parametrizing a permutationally invariant state has ten elements (including observables containing identity operators). In order to reach that quorum, we measure six different observables formed by projection operators (without identities). The list is given by $L = \{P_{X}\otimes P_{X},P_{Y}\otimes P_{Y},P_{Z}\otimes P_{Z},P_{X}\otimes P_{Y},P_{X}\otimes P_{Z},P_{Y}\otimes P_{Z}\}$, where $P_{X}$, $P_{Y}$ and $P_{Z}$ are the projection operators associated to eigenstates of Pauli matrices $X$, $Y$ and $Z$, with eigenvalue $+ 1$, respectively. Out of them, it is possible to compute the probabilities of those observables containing identities following a standard procedure. Notice that, for example, we include $P_{X}\otimes P_{Y}$, but not $P_{Y}\otimes P_{X}$. Since we are assuming that the state is permutationally invariant, there is no need to measure it (and a similar consideration holds in higher dimensions). Of course, deviations from the symmetry render this assumption false, and affect the quality of the estimation. Complete tomographic methods that do not consider the symmetry of the state need $9$ different observables (out of which one can compute the remaining observables that include identities, being $15$ in total).
%
%

For the case of three qubits (Figure \ref{f:3q_GHZ_parallel}), we find that there is a good agreement between the GIT method and the complete VQT. This indicates that for approximately symmetric states, and when noise levels are reasonable, the information obtained using the GIT methodology is equivalent to that of complete VQT. The same can be said about the relation between GIT and MaxLik. Also, since the fidelities are relatively high, taking into account the numerical simulations of section \ref{s:Numerical_Simulations}, one can expect that the GIT density operator is a good estimator of the real state (so the quantum processor succeeded in producing the expected states in most cases).

For the five and seven-qubit examples,
we only performed the GIT method obtaining the fidelities with respect to the target state, indicated in Figures \ref{f:5q_GHZ_Harmony} and \ref{f:7q_GHZ_Aria-1}. These results suggest a high performance of the Aria-1 quantum processor, since such a state is relatively challenging due to the size of the problem.

Let us now consider a new family of states, which is obtained by adding Hadamard gates to each qubit of Eq. \eqref{e:Parallel_phases}:

\begin{equation}\label{e:Parallel_twisted}
|\phi_{\theta}\rangle = \frac{1}{\sqrt{2^{N+1}}}\left(\sum_{x\in\{0,1\}^{N}} (1 + (-1)^{\oplus_{i}x_{i}}e^{i\theta})|x\rangle\right)
\end{equation}

\noindent where the sum $\oplus_{i}x_{i}$ is taken modulo $2$. A quick check shows that the states of Eq. \eqref{e:Parallel_twisted} are permutationally invariant. The corresponding circuit is displayed in Figure \ref{f:GHZ_parallel_twisted}.


\begin{figure}[H]
\centering
 \resizebox{\columnwidth}{!}{
\begin{quantikz}
\lstick[6]{N-qubits}  & \gate{H} &\gate{R_Z(\theta_n)}& \ctrl{1} &\qw&\qw &\qw  &\gate{H}&\qw\\
&\qw&\qw&  \targ{}  &\ctrl{1} & \qw & \qw   &\gate{H}&\qw      \\
&\qw&\qw&  \qw & \targ{} & \ctrl{1} & \qw&\gate{H}&\qw \\
&&&&& \vdots  &&& \\
&\qw&\qw&  \qw & \qw&  \targ{}  & \ctrl{1}&\gate{H}&\qw \\
&\qw&\qw&  \qw & \qw & \qw & \targ{}  &\gate{H}&\qw\\
\end{quantikz}}
\caption{Scheme of the circuits used to generated instances of the states given in  Eq. \eqref{e:Parallel_twisted}. For the three-qubit case, the angles are given by $\theta = 0, \frac{\pi}{14}, \frac{\pi}{7}, \frac{3\pi}{14}, \frac{2\pi}{7}, \frac{5\pi}{14}, \frac{3\pi}{7}, \frac{\pi}{2}$, and each of the three-qubit states are codified in qubits $0-2$, $3-5$, $6-8$, $9-11$, $12-14$, $15-17$, $18-20$, and $21-23$, respectively. For the six-qubit case, we used $\theta = 0, \frac{\pi}{6}, \frac{\pi}{3}, \frac{\pi}{2}$, and the six-qubit states were codified in qubits $0-5$, $6-11$, $12-17$, and $18-23$, respectively.}
\label{f:GHZ_parallel_twisted}
\end{figure}

In order to illustrate the experiment, we describe how the above circuits were implemented for three and six-qubit systems on Aria-1. For the three-qubit case, the angles are given by $\theta = 0, \frac{\pi}{14}, \frac{\pi}{7}, \frac{3\pi}{14}, \frac{2\pi}{7}, \frac{5\pi}{14}, \frac{3\pi}{7}, \frac{\pi}{2}$, and each of the three-qubit states are codified in qubits $0-2$, $3-5$, $6-8$, $9-11$, $12-14$, $15-17$, $18-20$, and $21-23$, respectively. For the six-qubit case, we used $\theta = 0, \frac{\pi}{6}, \frac{\pi}{3}, \frac{\pi}{2}$, and the six-qubit states were codified in qubits $0-5$, $6-11$, $12-17$, and $18-23$, respectively. For the three-qubit experiment we used $1000$ shots per measurement basis, while for the six-qubit experiment, we used $2500$ shots.

\begin{figure}[H]
\centering
\includegraphics[width=1\linewidth]{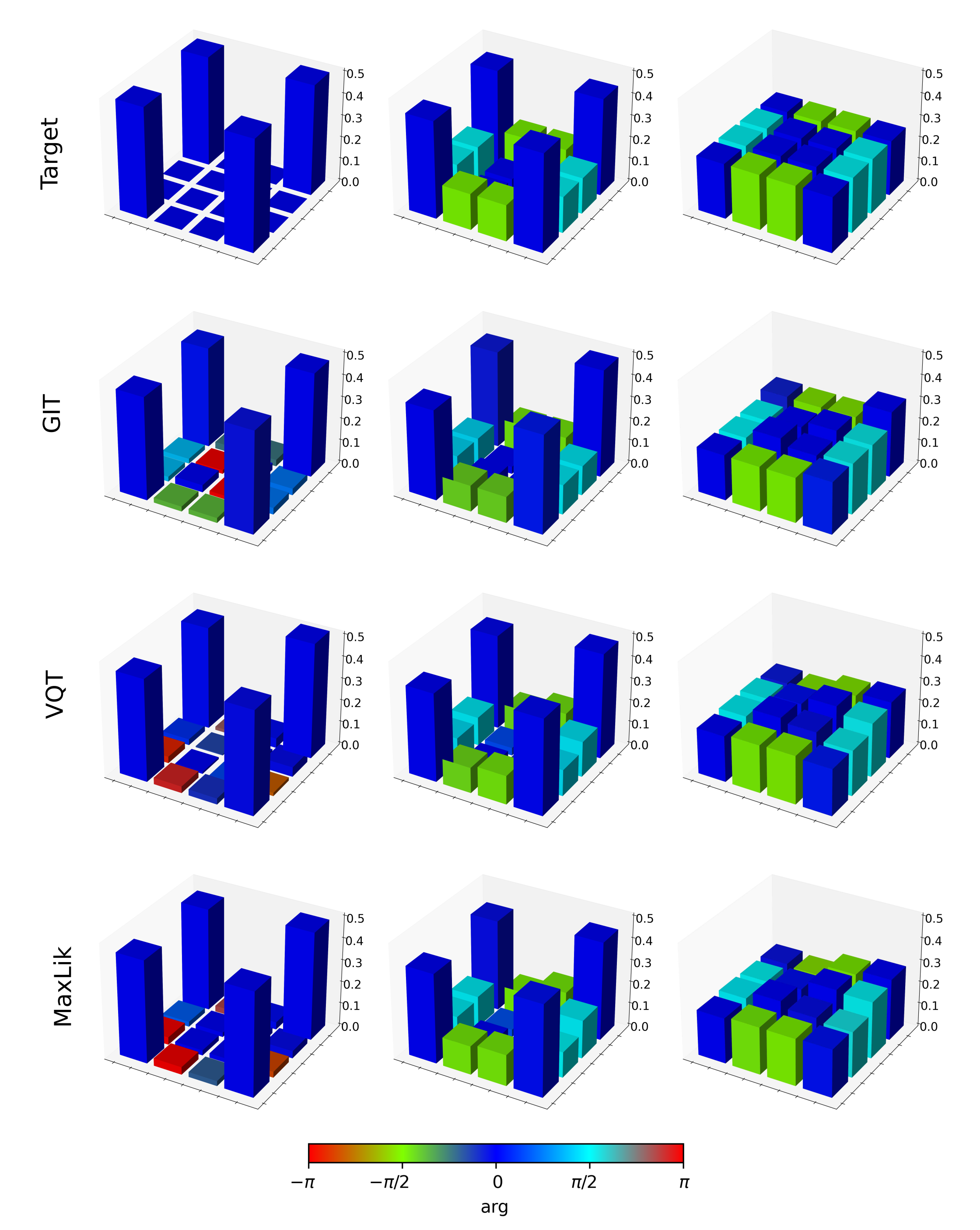}
\caption{Three instances of two-qubit states of the form of Eq. \eqref{e:Parallel_twisted} generated in Aria-1 quantum computer,
using $800$ shots for each measurement basis.
From left to right, the angles are given by $\theta = 0$, $\theta = 5\frac{\pi}{22}$ and $\frac{\pi}{2}$.
From top to bottom, we show the target state, and the estimated states using GIT, complete VQT, and MaxLik methods.
The mean fidelities obtained with the different estimation methods are:
GIT vs Target:  $0.96$,
cVQT vs Target:  $0.96$,
MaxLik vs Target:  $0.97$,
GIT vs cVQT: $0.99$,
GIT vs MaxLik:  $0.98$,
MaxLik vs cVQT:  $0.99$.}
\label{f:2q_paralel_twisted}
\end{figure}

\begin{figure}[H]
\centering
\includegraphics[width=1\linewidth]{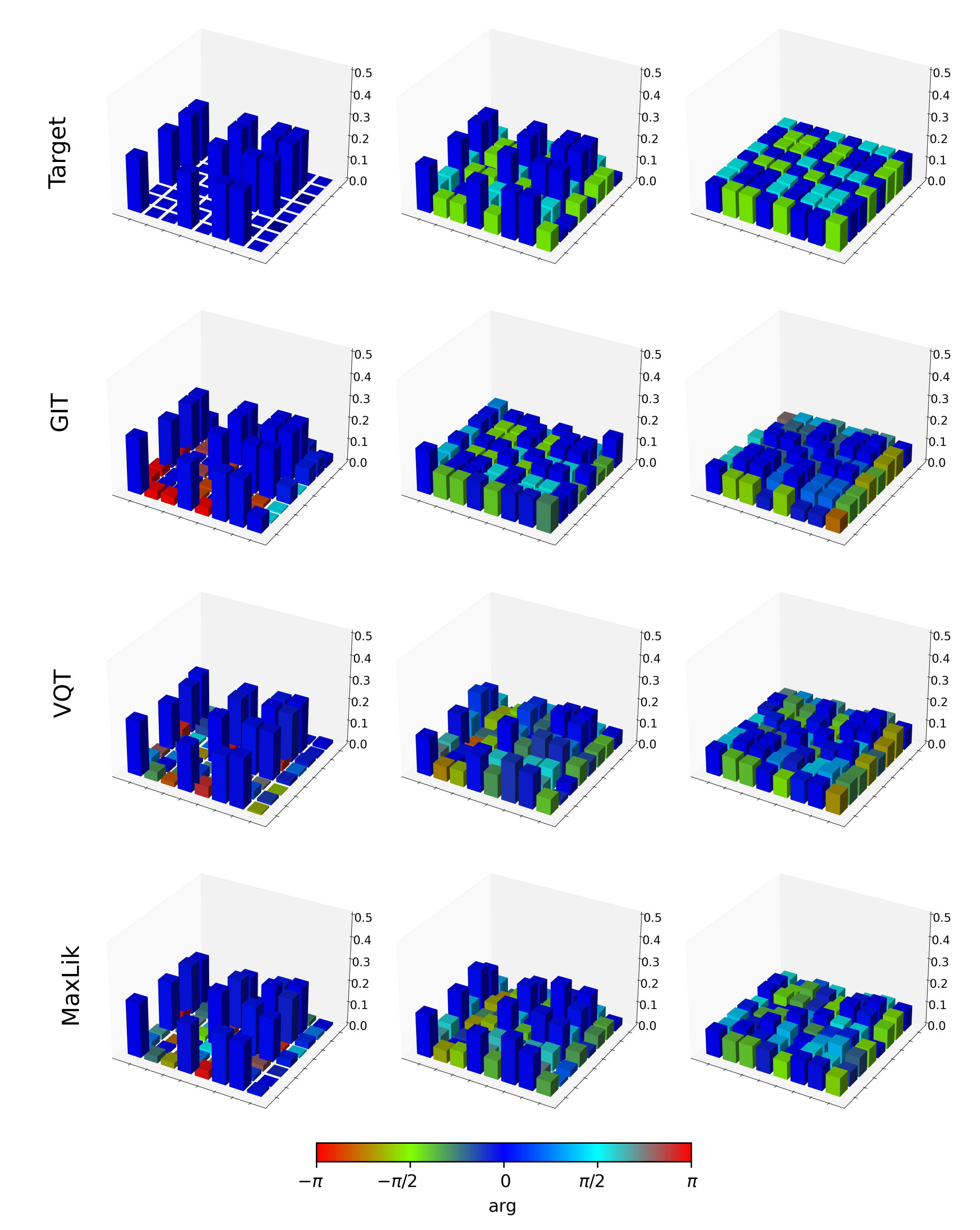}
\caption{Three instances of three-qubit states of the form of Eq. \eqref{e:Parallel_twisted} generated in Harmony quantum computer, using $1000$ shots for each measurement basis. From left to right, the angles are given by the angles are given by $\theta=0$, $\theta=\frac{\pi}{4}$, and $\theta=\frac{\pi}{2}$.
From top to bottom, we show the target state, and the estimated states using GIT, complete VQT, and MaxLik methods.
The mean fidelities obtained with the different estimation methods are:
GIT vs Target:  $0.86$
cVQT vs Target:  $0.90$,
MaxLik vs Target: $0.90$,
GIT vs cVQT: $0.92$,
GIT vs MaxLik:$0.89$,
cVQT vs MaxLik: $0.93$.}
\label{f:3q_paralel_twisted_Harmony}
\end{figure}

\begin{figure}[H]
\centering
\includegraphics[width=1\linewidth]{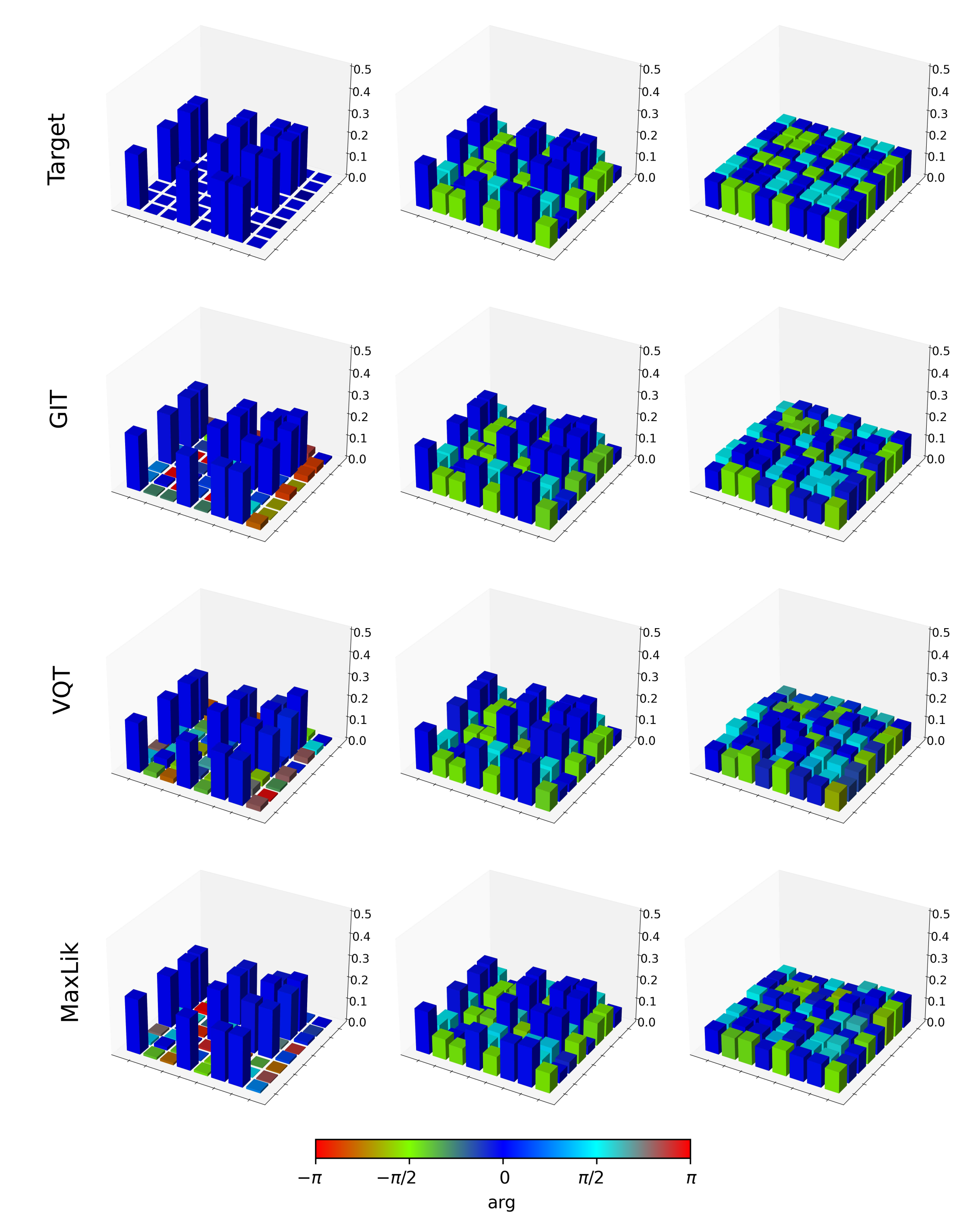}
\caption{Three instances of three-qubit states of the form of Eq. \eqref{e:Parallel_twisted} generated in
Aria-1 quantum computer,
using $1000$ shots for each measurement basis.
From left to right, the angles are given by $\theta =$ $0$, $\frac{3\pi}{14}$, and $\frac{\pi}{2}$.
From top to bottom, we show the target state, and the estimated states using GIT, complete VQT, and MaxLik methods.
The mean fidelities obtained with the different estimation methods are:
GIT vs Target:  $0.94$,
cVQT vs Target:  $0.94$,
MaxLik vs Target:  $0.95$,
GIT vs cVQT:  $0.95$
GIT vs MaxLik:  $0.94$,
cVQT vs MaxLik:  $0.96$.}
\label{f:3q_paralel_twisted_Aria}
\end{figure}

\begin{figure}[H]
\centering
\includegraphics[width=1\linewidth]{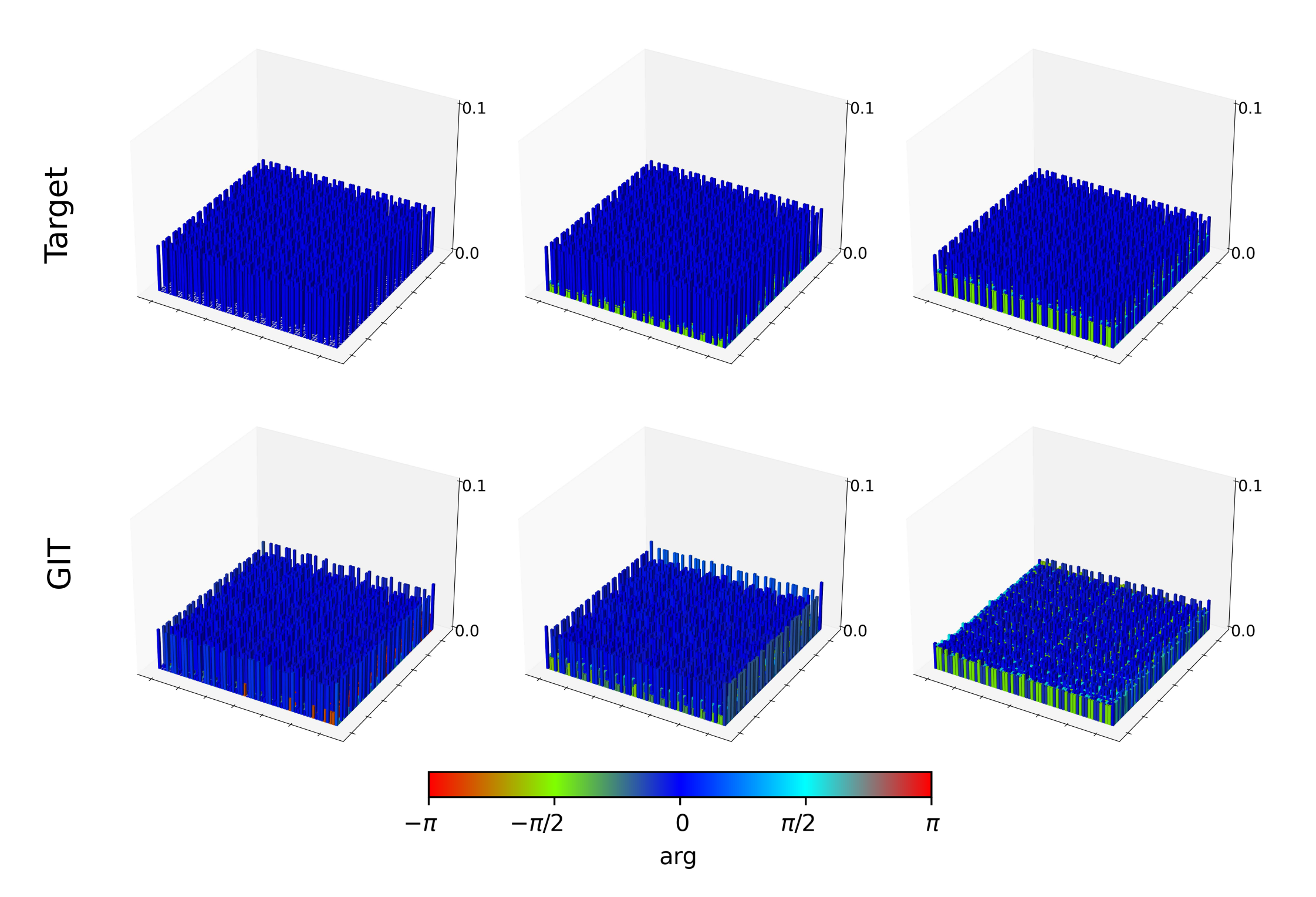}
\caption{
Three instances of six-qubit states generated in Aria-1 quantum processor using the circuit of Figure \ref{f:GHZ_parallel_twisted}, with $2500$ shots for each measurement basis.
From left to right, the angles are given by $\theta = 0$, $\frac{\pi}{6}$ and $\frac{\pi}{2}$.
From top to bottom, we show the target state, and the estimated state using GIT method.
The mean fidelities between the GIT estimation and the target state are given by $0.85$, $0.89$, and $0.90$, respectively.}
\label{f:6q_paralel_twisted}
\end{figure}

\begin{figure}[H]
\centering
\includegraphics[scale=0.3]{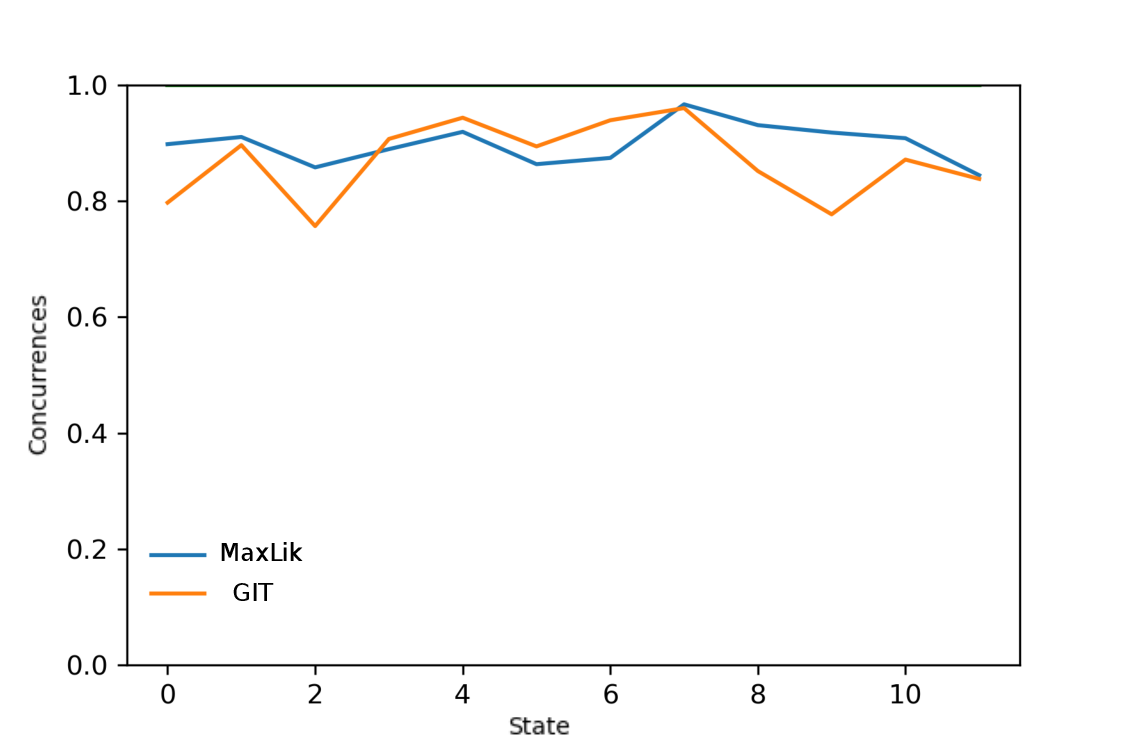}
\caption{Concurrences of all the two-qubit states of the form given by Eq. \eqref{e:Parallel_twisted} generated in Aria-1 (twelve in total; some instances are shown in Figure \ref{f:2q_paralel_twisted}). The mean values of the absolute value of the differences between GIT vs Target, GIT vs MaxLik, and MaxLik vs Target are $0.13$, $0.05$ and $0.10$
, respectively. This implies a $5\%$ difference between the concurrences obtained using MaxLik and the GIT estimation.}
\label{f:2q_GHZ_concurrences_26_06}
\end{figure}

In Figures \ref{f:2q_paralel_twisted},
\ref{f:3q_paralel_twisted_Harmony},
\ref{f:3q_paralel_twisted_Aria}, and \ref{f:6q_paralel_twisted} we show instances of the states given in Eq. \eqref{e:Parallel_twisted} for two, three and six qubits, respectively. For two and three qubits, we obtained complete tomographic information, in order to make the comparison. The fidelities are quite high, taking into account that the states are highly entangled and superposed. The concurrences for the two-qubit states are displayed in Figure \ref{f:2q_GHZ_concurrences_26_06}.

\subsection{Werner states of two and four qubits}

The simplest example of Werner states is given by the two-qubit case, which can be parametrized as:

\begin{equation}\label{e:Werner_2q_equation}
W_{p} = \left(\frac{1-p}{4}\right)\sigma_{0}\otimes\sigma_{0}+p|\psi^{-}\rangle\langle\psi^{-}|
\end{equation}

\noindent These states have the following symmetry:
$(U\otimes U)W_{p}(U^{\dagger}\otimes U^{\dagger}) = W_{p}$, with $U$
an arbitrary unitary operator. Werner states have a straightforward generalization for a higher number of qubits, but in that cases exact parametrization formulas are more challenging (see Ref.~\cite{Eggeling_PhD} and Ref.~\cite{Eggeling2001}). Using the GIT methodology it is possible to find the basis for Werner states numerically.

\begin{figure}[H]
\centering
\resizebox{\columnwidth}{!}{
     \begin{tikzcd}
    & \gate{R_Y(\theta_{a})} &\ctrl{1}&  \gate{R_Y(\theta_{b})} &\ctrl{2}  & \gate{H} &\ctrl{1}   &\qw \rstick[2]{Werner\\ State}\\
    &\qw& \targ{} & \gate{R_Y(\theta_{b})} & \qw  &\ctrl{2}  &  \targ{} &  \qw  \\
    &\qw&\qw&  \qw &\targ{}  & \qw   & \qw  \rstick{\textbf{]}}\\
     &\qw&\qw&  \qw & \qw &  \targ{}   &\qw  \rstick{\textbf{]}}
    \end{tikzcd}}
\caption{Scheme of the circuits used to generated
instances of
two-qubit Werner states (first two qubits) using two ancillas (last two qubits).}
\label{f:Werner_Circuit}
\end{figure}
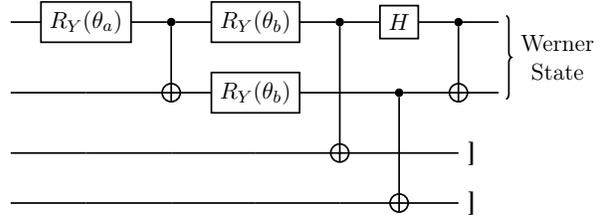

In order to build this type of states, we use the circuit of Figure \ref{f:Werner_Circuit}, following the proposal described in Ref.~\cite{Werner_in_IBMQ}. The values chosen for $\theta_a$ and $\theta_b$ are shown in the Table \ref{tab:2q_Werner_paralel}.

\begin{table}[H]
    \centering
\begin{tabular}{ c c c }
p     & $\theta_{a}$ & $\theta_{b}$ \\
\hline
1.0    & 0.00 & 3.14\\
0.76 & 0.33    & 2.50\\
0.40 & 0.27    & 2.00\\
\end{tabular}
    \caption{Values of circuit angles (Figure \ref{f:Werner_Circuit}) for quantum states of Figure  \ref{f:2q_Werner_paralel}. From top to bottom, each row correspond to the quantum state from left to right of Figure \ref{f:2q_Werner_paralel}.}
    \label{tab:2q_Werner_paralel}
\end{table}

\begin{figure}[H]
\centering
\includegraphics[width=1\linewidth]{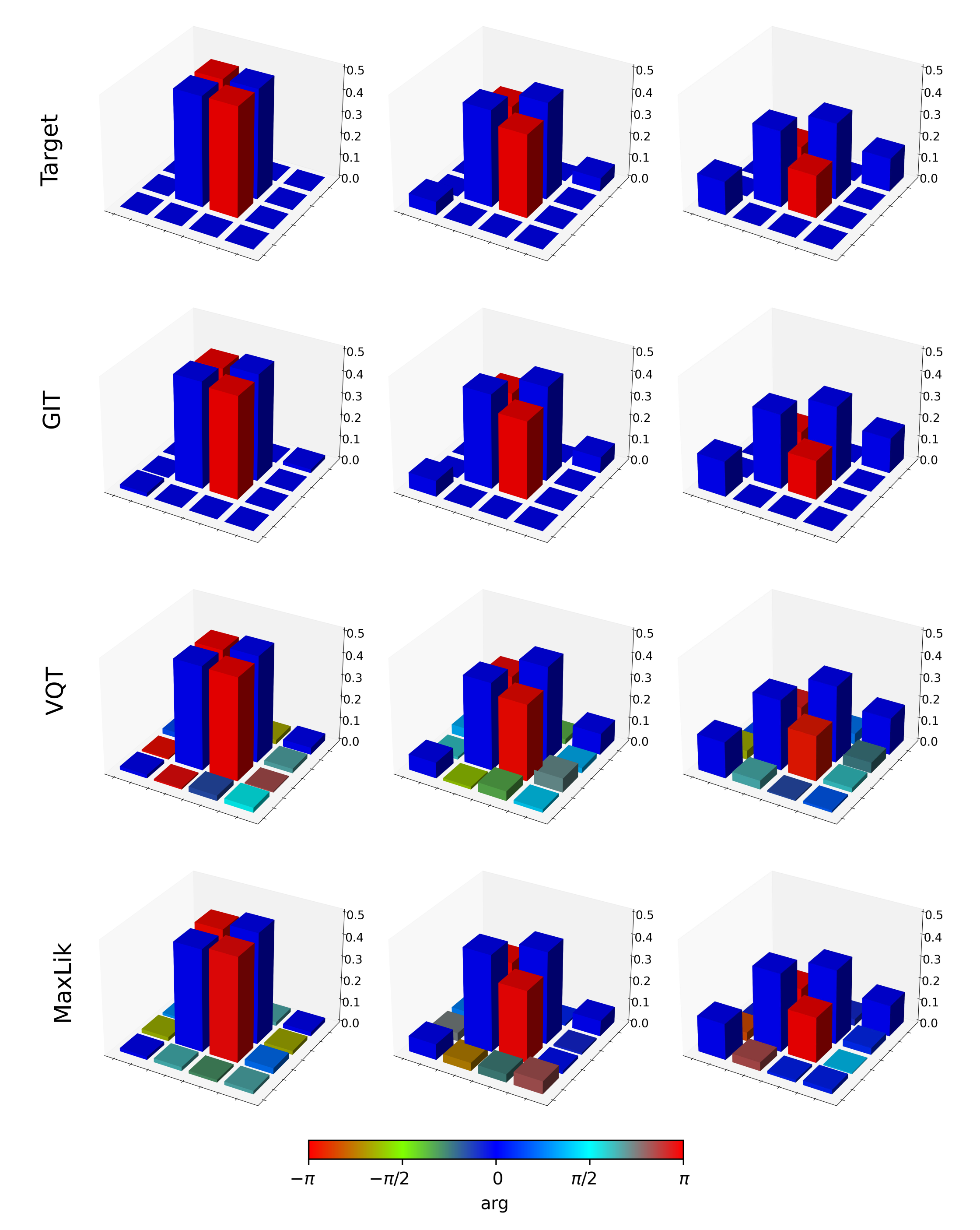}
\caption{Three instances of two-qubit states generated
in Harmony quantum processor using the circuit of Figure \ref{f:Werner_Circuit}, with 300 shots for each basis measurement. The values of the parameter $p$ for each state are displayed in Table \ref{tab:2q_Werner_paralel}.
From top to bottom, we show the target state, and the estimated states using GIT, complete VQT, and MaxLik methods.
The mean fidelities obtained with the different estimation methods are:
GIT vs Target: $0.99$
cVQT vs Target: $0.98$,
MaxLik vs Target: $0.98$,
GIT vs cVQT:  $0.99$,
GIT vs MaxLik: $0.98$,
cVQT vs MaxLik:  $0.98$.}
\label{f:2q_Werner_paralel}
\end{figure}

\begin{figure}[H]
\centering
\includegraphics[scale=0.3]{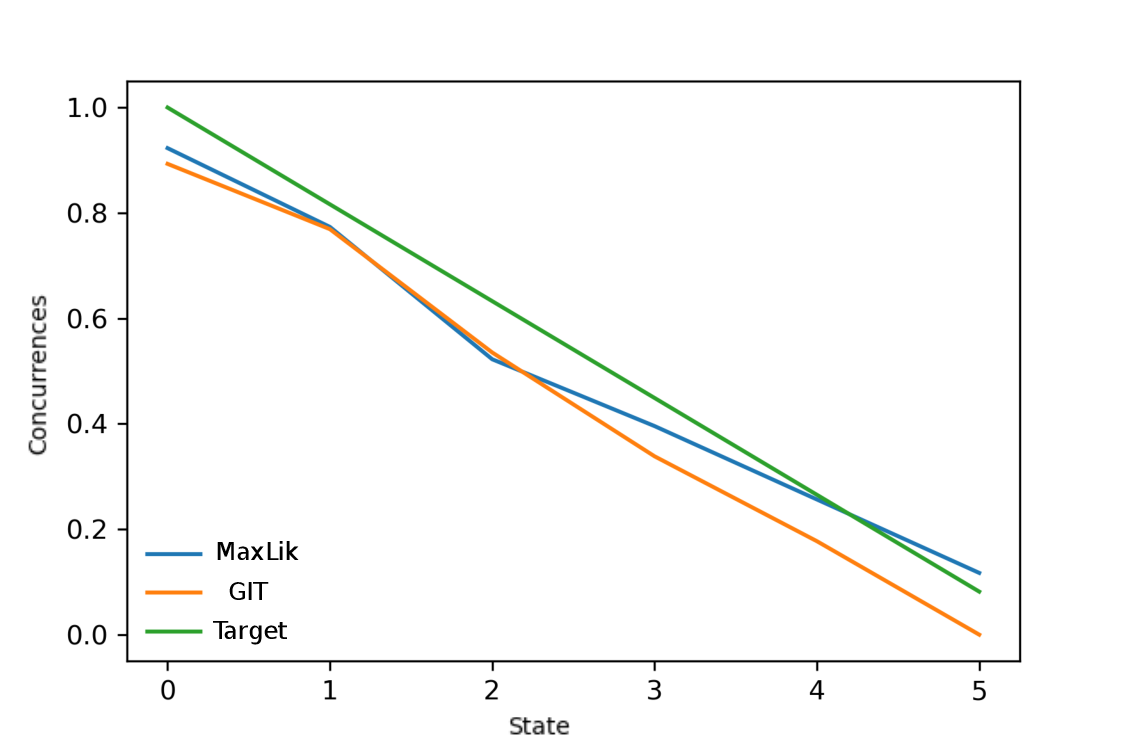}
\caption{Comparison of the concurrences of the target state, GIT estimation,
and MaxLik estimation for all Werner states. The mean values of the absolute value of the differences between GIT vs Target, GIT vs MaxLik, and MaxLik vs Target are $0.09$, $0.05$ and $0.06$
, respectively.}
\label{f:2q_Werner_concurrences}
\end{figure}

In Figure \ref{f:2q_Werner_paralel} we show the results of generating three two-qubit Werner states in Aria-1 quantum computer, using the values of the  circuit angles given in the Table \ref{tab:2q_Werner_paralel}. The mean fidelities between the target state and the GIT estimation are all above $0.97$. In Figure \ref{f:2q_Werner_concurrences}, we show the concurrences for the target state, GIT estimation, and MaxLik estimation. We see a high degree of agreement between the values obtained with GIT and MaxLik ($5\%$ difference in the mean).

In order to produce a four-qubit Werner state we use the circuit of Figure \ref{f:Werner_Circuit}, in parallel, on different sets of qubits. Such a procedure creates the tensor product of two-qubit Werner states. By adjusting the parameters, one can change the degree of entanglement of the global state. The corresponding Werner states (and thus, their measurements in different basis) are implemented in qubits $0-3$. It is important to take into account the following. From the point of view of the agent who generates the states, these are relatively simple, since they are a simple product of two (two-qubit) Werner states. But, from the point of view of the agent who needs to detect the states, if the only promise is that the otherwise unknown state has Werner symmetry, the task is challenging. First, it could be a highly mixed state. Thus, a full tomography would imply performing $3^{4} = 81$ different experimental setups. Instead, in our example we have only measured in $15$ different basis (the Werner variety for four qubits has $14$ parameters). On the other hand, the states might not necessarily be permutationally invariant. Thus, the methodology used in Ref.~\cite{PermutationallyInvariantQT} cannot be applied here. Taking into account these challenges, we can see that the performance of the GIT method is quite promising, since the estimation is reasonably good and allows the user to save a lot of resources.

\begin{figure}[H]
\centering
\includegraphics[width=1\linewidth]{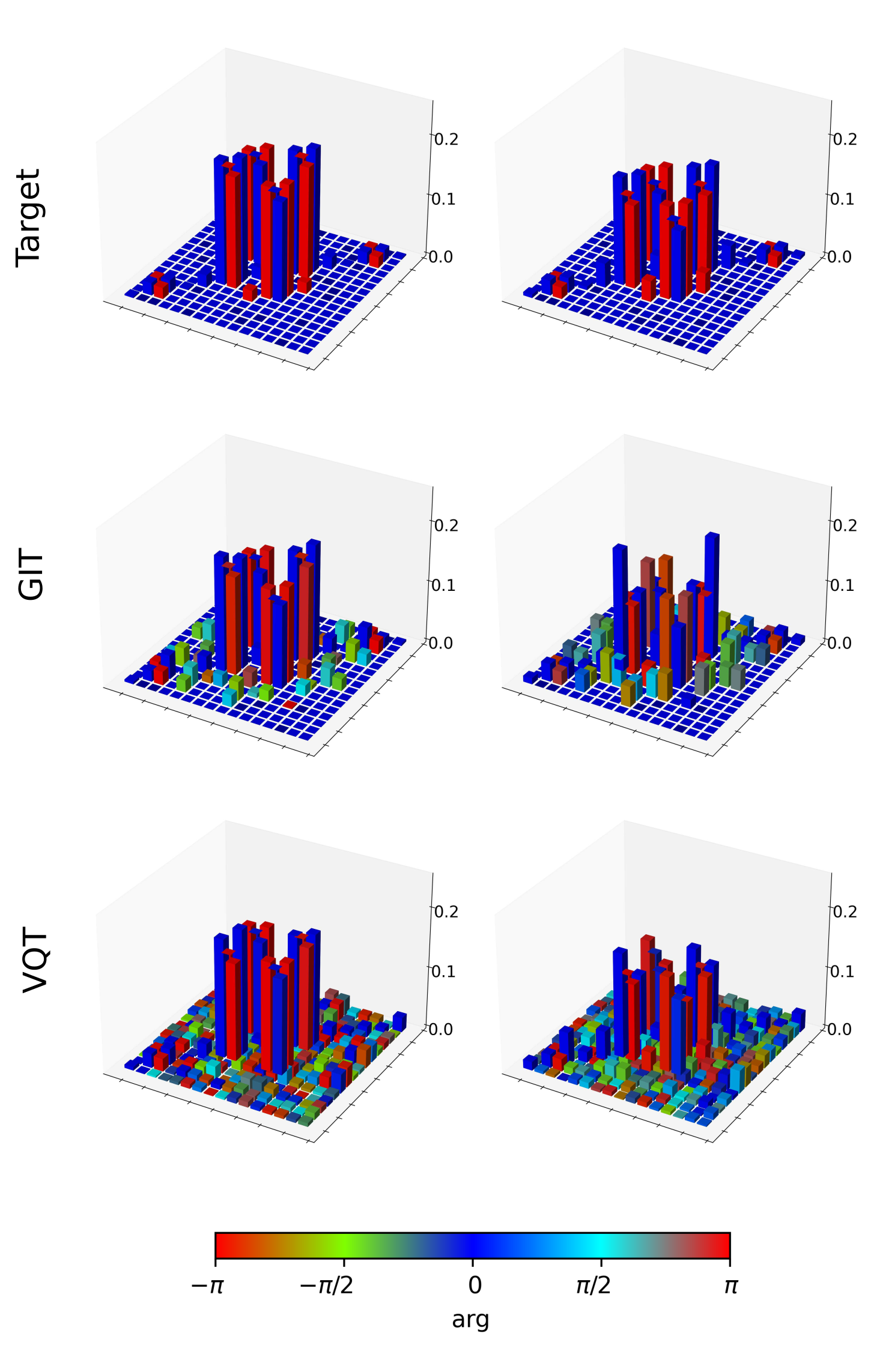}
\caption{Target state (top), complete VQT estimation (center) and GIT estimation (botton) of a four-qubit Werner state generated in Harmony quantum computer with $1000$ shots. In the left column we show the results for $W_{0.76}\otimes W_{0.63}$, and in the right those for $W_{0.81}\otimes W_{0.81}$. The angles used for generating those states are displayed in Table \ref{f:Harmony-4q-Werner-05-06-2023}.
The obtained fidelities are as follows: GIT vs Target $0.93$, $0.90$, GIT vs cVQT: $0.88$, $0.82$, and cVQT vs Target: $0.92$, $0.87$.}
\label{f:Harmony-4q-Werner-05-06-2023}
\end{figure}

Here we consider two experiments with four-qubit Werner states. Using two circuits, a four-qubit Werner state can be generated as a tensor product, $W_4=W_p\otimes W_{p'}$.  The results are shown in Figure \ref{f:Harmony-4q-Werner-05-06-2023}.

\begin{table}[]
    \centering
\begin{tabular}{ c c c }
 p     & $\theta_{a}$ & $\theta_{b}$ \\
 \hline
0.76 & 0.33 & 2.50\\
0.63 & 0.34 & 2.31\\
0.81 & 0.31 & 2.60

\end{tabular}
    \caption{Values of circuit angles (Fig. \ref{f:Werner_Circuit}) for quantum states of Fig. \ref{f:Harmony-4q-Werner-05-06-2023}. The values of the first two rows are used to generate, as a tensor product, the Werner states of the left columns. In the same way, the values in row three are used twice to obtain the density matrices in the right column.}
\end{table}

\subsection{Varying the number of observables with a MaxEnt-like estimation}\label{s:Varying_Observables}

As we remarked in the introduction, one of the main advantages of the MaxEnt estimation (and also of VQT) is that it is not necessary to have complete information about the system for the problem to converge and obtain an estimation. The same applies for the GIT technique, with regard to a new quorum defined by the number of parameters appearing in Eq. \eqref{e:Parametrization}. These techniques yield an estimation that is the least biased with regard to the available information. In this section we use some of the previous experiments to describe how the estimation varies with the number of measured observables. The reminder plots in this section should be compared to those of Ref.~\cite{tielas2021performance} (also with those of \cite{Goncalves-2013-MaxEntTomography}).

\begin{figure}[H]
\centering
\includegraphics[width=1\linewidth]{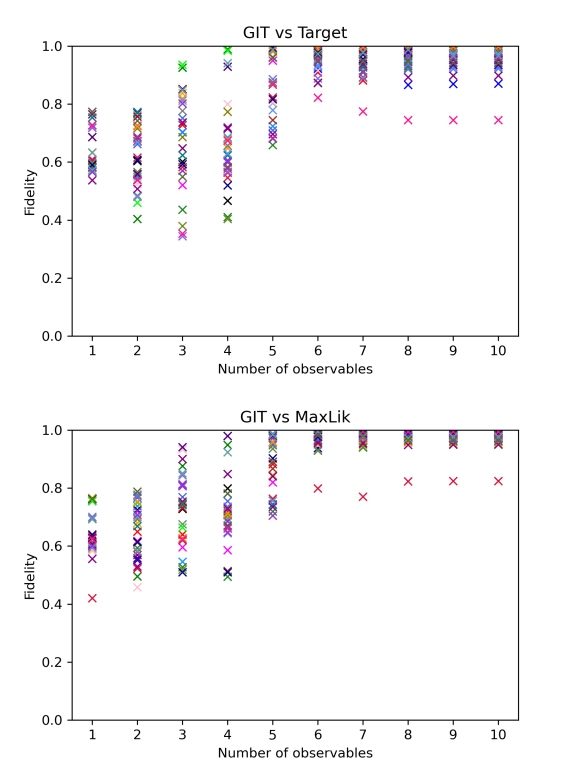}
\caption{ Fidelities of all two-qubit permutationally invariant states with respect to the target state (top) and the MaxLik state (bottom). For two-qubit states, there are $6$ different measured observables, from which we compute the remaining four observables with identities (so there are $10$ in total). The parametrization basis has ten elements, so when we use ten observables, quorum is reached with regard to that basis.}
\label{f:2q_permutational_target_variable_observables}
\end{figure}

\begin{figure}[H]
\centering
\includegraphics[width=1\linewidth]{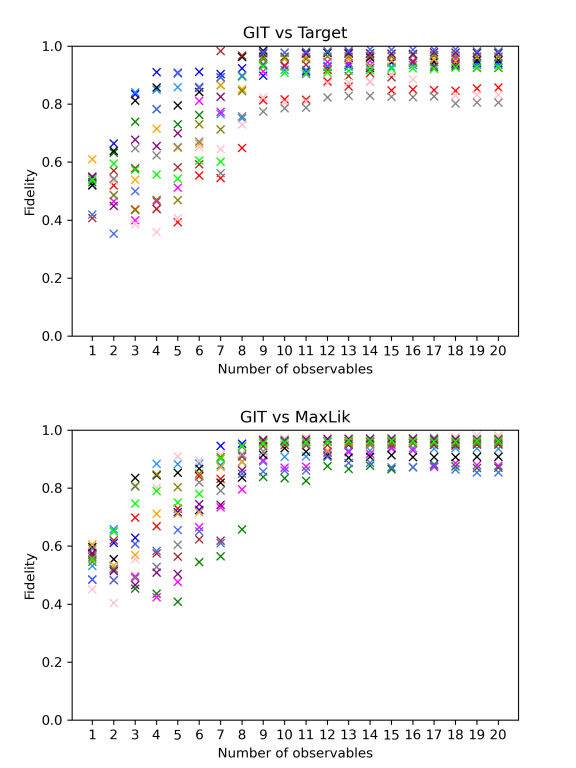}
\caption{Fidelities of all three-qubit permutationally invariant states with respect to the target state (top) and the MaxLik state (bottom). For three-qubit states, there are $10$ different measured observables, from which we compute the remaining $10$ observables with identities (so there are $20$ in total).}
\label{f:3q_permutational_target_variable_observables}
\end{figure}

\begin{figure}[H]
\centering
\includegraphics[width=1\linewidth]{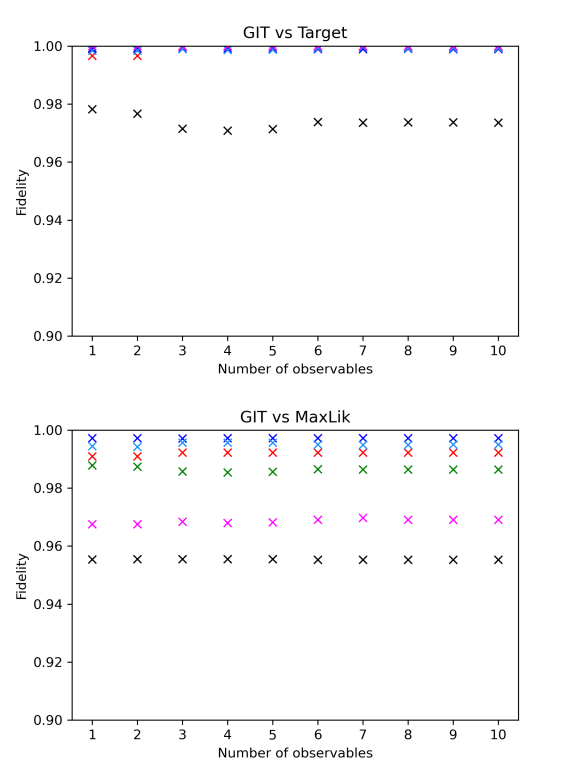}
\caption{Fidelities of all two-qubit Werner states with respect to the target state (top) and MaxLik state (bottom). For this case, there are $6$ projections and the parametrization basis has only two elements. Therefore, information converges very fast to maximal (and the reminder observables are not necessary).}
\label{f:2q_Werner_target_variable_observables}
\end{figure}

In Figure \ref{f:2q_permutational_target_variable_observables} we show the fidelities of the states of Figure \ref{f:2q_paralel_twisted} with respect to the target states and MaxLik states, obtained for different numbers of measured observables. Notice that for a two-qubit permutationally invariant state one needs to measure $6$ different observables. All the observables containing identity operators can be obtained from the measured observables. There are $10$ of them in total. Consequently, in Figure \ref{f:2q_permutational_target_variable_observables}, the number of observables ranges from $1$ to $10$. Figure \ref{f:3q_permutational_target_variable_observables} shows the fidelities with respect to the target and MaxLik estimation for all the three qubits experiments of permutationally invariant states we did (some of the tomographies are displayed in Figures \ref{f:3q_GHZ_parallel}, \ref{f:3q_paralel_twisted_Harmony}, and \ref{f:3q_paralel_twisted_Aria}). The same is done in Figure \ref{f:2q_Werner_target_variable_observables} for the Werner states of Figure \ref{f:2q_Werner_paralel}.

\begin{figure}[H]
\centering
\includegraphics[width=1\linewidth]{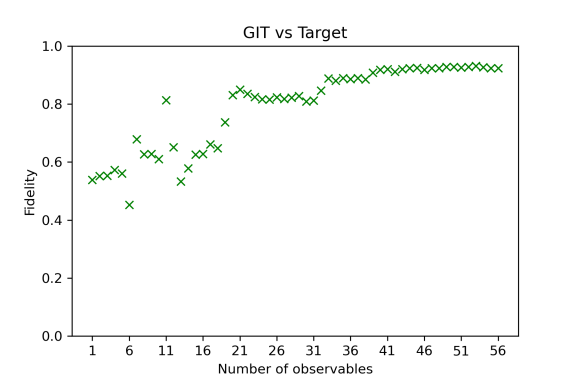}
\caption{Five-qubit GHZ state statistics obtained in Harmony quantum computer. For five qubits, the parametrization basis has $20$ elements.
We measure in $10$ different basis, out which we compute $20$ projections containing identities.}
\label{f:5q_GHZ_variable_observables}
\end{figure}

\begin{figure}[H]
\centering
\includegraphics[width=1\linewidth]{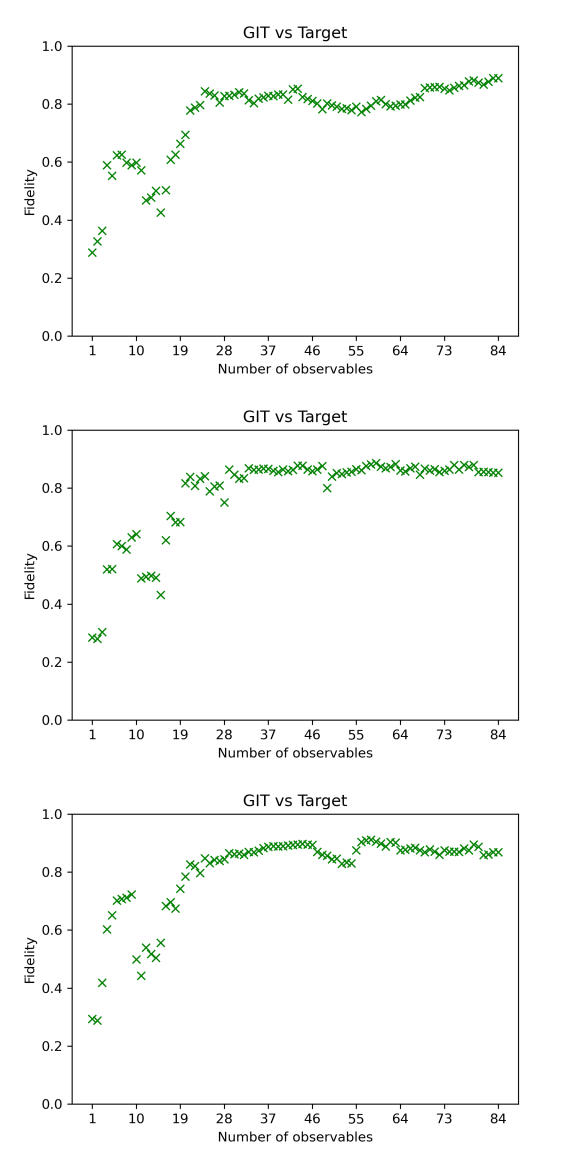}
\caption{From top to bottom, three examples of six-qubit states of the form of Eq. \eqref{e:Parallel_twisted}, with phases equal to $\frac{\pi}{6}$, $\frac{\pi}{3}$ and $\frac{\pi}{2}$ (see Figure \ref{f:6q_paralel_twisted}). For six qubits, the permutational basis has $84$ elements. We measure in $28$ different basis, out which we compute $84$ observables with identities.}
\label{f:6q_GHZ_twisted_variable_observables}
\end{figure}

\begin{figure}[H]
\centering
\includegraphics[width=1\linewidth]{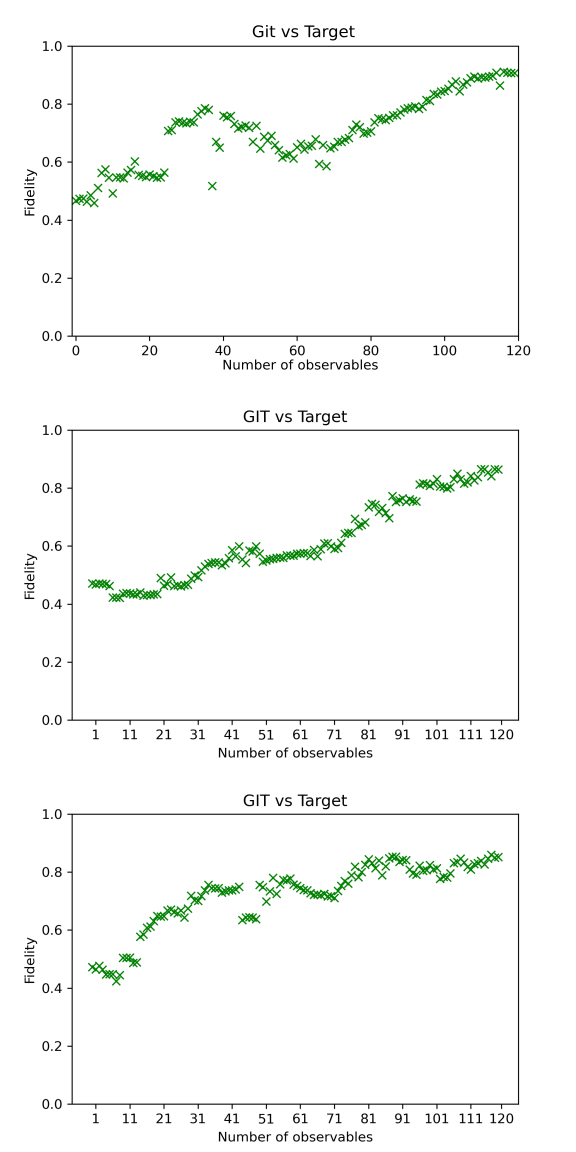}
\caption{For seven qubits, the parametrization basis has 120 elements. We perform measurements in $36$ different basis, from which we compute $120$ observables with identities.}
\label{f:7q_GHZ_variable_observables}
\end{figure}

Notice that, while quorum with respect to the parametrization basis is reached at $10$ and $20$ observables for two and three-qubit permutationally invariant states respectively, the fidelities for many states stabilize in a higher value for a fewer number of observables. The situation is more extreme for the case of two-qubit Werner states. This remarkable fact becomes clearer for a higher number of qubits. In Figures \ref{f:5q_GHZ_variable_observables}, \ref{f:6q_GHZ_twisted_variable_observables}, and \ref{f:7q_GHZ_variable_observables}, we plot the fidelities with respect to the target states of a five-qubit GHZ states, a six-qubit state of the form of Eq. \eqref{e:Parallel_twisted}, and a seven-qubit GHZ state, respectively, ranging the number of observables from one to the maximal number with regard to the quorum determined by the parametrization basis.

\begin{figure}[H]
\centering
\includegraphics[width=1\linewidth]{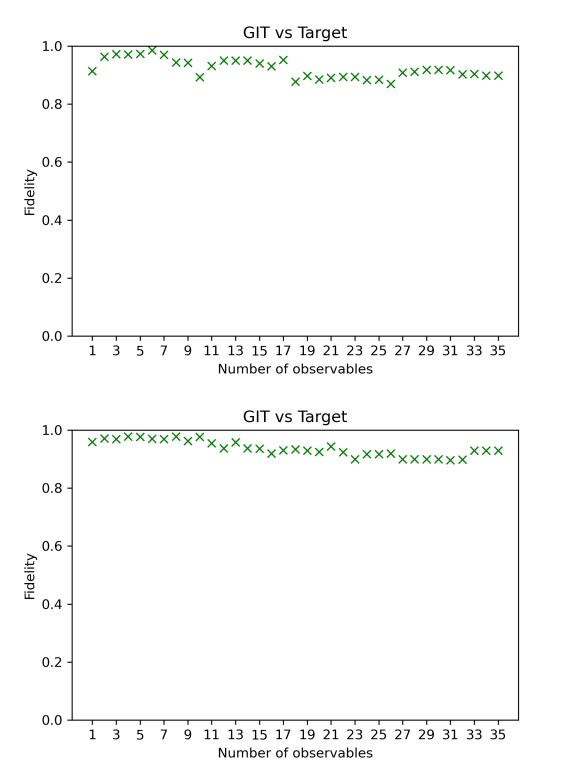}
\caption{For four qubits, the parametrization basis of a Werner state has $14$ elements. Thus, using suitably chosen proyectors, one can recover a great part of the state. We performed measurements in $15$ different basis, from which we computed $35$ observables with identities. Notice that, for this case, it is not necessary to appeal to all observables.}
\label{f:4q_Werner_variable_observables}
\end{figure}

\section{Discussion}\label{s:Conclusions}

In this work we have studied the application of the GIT methodology in quantum processors. We have performed both, numerical experiments and implementations in IonQ quantum computers. The results obtained indicate that:

\begin{itemize}
    \item The GIT estimation is in good agreement with the complete VQT estimation. For two and three qubits we have also found a good agreement between GIT, VQT, and MaxLik.
    \item The GIT estimation allows for a substantial reduction of the resources needed to perform the estimation (both, experimental and computational).
    \item The effect of noise (in preparation and measurement) and numerical instabilities become critical for complete tomographic methods when the number of qubits scales up. Thus, there seems to be a trade-off between acquiring less information and making a guess (with the GIT method) on one hand, and acquiring a lot of information, but accumulating a lot of errors on the other (complete VQT).
    \item From the point of view of the end user, GIT can be used as a fast and cheap method to assess whether a quantum device is capable of generating certain symmetric states. Since the set of symmetric states with different symmetries is very rich in quantum resources (entanglement, non-locality and contextuality, as is the case for GHZ states), this methodology can be used to assess the quantum device as a whole.
    \item In particular, GIT yields very good results for the task of estimating Werner states, which can be highly mixed, and highly entangled at the same time.
    \item For some states it is possible to obtain a reasonable performance even when the  number of measured observables is far from the quorum dictated by the parametrization given in Eq. \eqref{e:Parametrization}. These results indicate that there is a clear advantage in using GIT method with regard previous options in the literature. It is important to remark that the results presented in Section \ref{s:Varying_Observables} can be considered as a experimental verification of the theoretical studies presented in Ref.~\cite{tielas2021performance} and Ref.~\cite{Goncalves-2013-MaxEntTomography}.
\end{itemize}

Overall, the numerical simulation and experiments in quantum processors indicate that the GIT technique is a good alternative to have at hand in many situations of interest. In future works, we will address the problem of scaling up the number of qubits, and improving the quality of the numerical postprocessing. The main goal of our research project is to develop a ready-to-use platform for estimating symmetric quantum states for everyday quantum physics applications.

\section*{Acknolwedgements}\label{s:Aknol}

This research was funded by the project INNOVA-CONICET - AWS: \textit{Algorithms for estimation and generation of symmetric quantum states}. The authors wish to thank Sebastian Bassi and Virginia Gonz\'{a}lez (from Toyoko LLC) for their technical assistance during the use of the AWS Braket API. The authors also thank Hilary Foster and Vanessa Hernández Mateos of AWS Braket for the overall assistance during the execution of the project.


\providecommand{\noopsort}[1]{}\providecommand{\singleletter}[1]{#1}%

\onecolumn




\end{document}